\documentstyle[11pt,aasms4]{article}
\tighten

\newcommand{\lta}{\stackrel{<}{\sim}}
\newcommand{\gta}{\stackrel{>}{\sim}}

\newcommand{\lbar}{\overline}

\lefthead{Hubeny \& Hubeny}
\righthead{Non-LTE models of AGN disks}

\begin{document}

\title{Non-LTE Models and Theoretical Spectra of Accretion Disks in
Active Galactic Nuclei II. Vertical Structure of the Disk}

\author{ Ivan Hubeny\altaffilmark{1}}
\affil{AURA/NOAO,
       NASA Goddard Space Flight Center, Code 681, Greenbelt, MD 20771} 
\and
\author{ Veronika Hubeny\altaffilmark{2}}
\affil{ Department of Physics, 
University of California, Santa Barbara, CA 93106}

\altaffiltext{1}{hubeny@tlusty.gsfc.nasa.gov}
\altaffiltext{2}{veronika@physics.ucsb.edu}

\begin{abstract}
We have calculated several representative models of vertical structure
of an accretion disk around a supermassive Kerr black hole. The interaction
of radiation and matter is treated self-consistently, taking into account
departures from LTE for calculating both the disk structure and the
radiation field. The structural equations are described in detail, and
various approximations are discussed.
We have demonstrated that departures from LTE are very important for
determining the disk structure, even at the midplane, as well as the 
emergent radiation, particularly for
hot and electron--scattering--dominated disks. We have shown that at least
for the disk parameters studied in this paper, NLTE effects tend to
reduce the value of the Lyman jump with respect to the
LTE predictions, regardless whether LTE predicts an emission or
absorption jump.
We have studied the effects of various values of viscosity on the model
structure and predicted spectral energy distribution.
The viscosity is parameterized through a
parameter $\alpha_0$ which describes the vertically-averaged viscous stress,
two power-law exponents $\zeta_0$ and $\zeta_1$, and the division point
$m_{\rm d}$ between these two forms. The disk structure and emergent radiation 
is sensitive mainly to the values of $\alpha_0$, while the other parameters
influence the disk structure to a much lesser extent. However, although
the detailed shape of the predicted spectrum is sensitive to adopted value of
$\alpha_0$, the overall appearance of the spectrum is quite
similar.

%
%

\end{abstract}


\clearpage

%
%

\section{INTRODUCTION}

Accretion disks around massive black holes have long been the most
popular candidates for providing the
ultraviolet and soft X-ray flux observed in Active Galactic Nuclei (AGN).
Observational evidence is mainly
based on the `big blue bumps' seen in the UV (e.g. Shields 1978; 
Malkan \& Sargent 1982). However, despite its attractiveness this model
faces a number of problems when confronted with multi-wavelength observations.
The current situation has been recently summarized by Koratkar (1998)
from the observational point of view, and by Blaes (1998) from the
theoretical viewpoint. The most pressing problems of the theoretical models
are a near absence of observed Lyman discontinuity (first pointed out by 
Antonucci, Kinney, \& Ford 1989), the UV/EUV continuum polarization
(e.g., Antonucci 1992, Blaes 1998 and references therein), 
the overall continuum spectral energy distribution (e.g. Laor 1990), 
and phased optical/UV variability (e.g., Alloin et al 1985).

What is clearly needed is a self-consistent model which would explain all
the observed features. Such a goal is still far away, but we feel that the 
first step towards it is to answer the fundamental question: In view of all
current problems, is the accretion disk paradigm still a viable model?
In other words, are the observed phenomena truly inconsistent with
the accretion disk picture in general, or is the lack of agreement rather
a result of inaccuracies or even inconsistencies in the computed models
or in the current modeling techniques?

Therefore, we have embarked on a systematic study of these questions.
We recognize that there are many possible sources of inconsistencies
in the AGN accretion disk modeling. There are essentially two types of
problems. First, there are basic physical uncertainties. Among them,
the most important is our ignorance of the basic physics of viscous
energy dissipation in the AGN disks, so that we are left with a necessity
to employ certain {\it ad hoc}
parameters. Although this is certainly a viable approach when nothing else
is currently available, we should bear in mind that corresponding models will
lack predictive power---we may explain what we see, but we cannot predict it.
In any case, when adopting a model based on some chosen 
set of parameters, one at least has to study carefully the influence of
these parameters on the computed model.
Other fundamental physical problems are uncertainties of the effects
of other structures forming the AGN complex onto the accretion disk;
for instance an irradiation of the disk from external sources. Again,
in the absence of any detailed theory, we are usually left with a necessity to
parameterize these effects.

Second type of problems concerns the degree of approximation used
in the actual modeling procedure. A particularly important class of
such approximations deals with the description of interaction of radiation 
and matter in the disk. It should be emphasized that AGN disks, like
atmospheres of hot stars, are typical examples of a medium where radiation
is not only a {\em probe} of the physical state, but in fact a crucial
{\em constituent}. In other words, radiation not only carries an information
about the medium,
it in fact {\em determines} its structure. Consequently, a treatment of this
interaction is in a sense the very gist of the problem. We should
therefore study the influence of various approximations, such as the degree
of equilibrium assumed, or, more specifically, the extent to which the local 
thermodynamic equilibrium (LTE) applies, and 
the completeness of considered
opacity and emissivity sources on the computed model, etc.

In the previous paper (Hubeny \& Hubeny 1997 -- hereafter called Paper I), 
we have presented some representative self-consistent, 
non-LTE models of the vertical structure of AGN accretion disks. 
The basic aim of that study was to investigate the differences in the predicted 
spectrum between this and previous approaches  (Sun \& Malkan 1989; 
Laor \& Netzer, 1989; Ross, Fabian, \& Mineshige 1992;
Wehrse et al. 1993; St\"orzer and Hauschildt 1994; Coleman 1994; 
Shields \& Coleman 1994;
Blaes \& Agol 1996; D\"orrer et al. 1996). The emphasis was to
clarify the role of departures from LTE and to study the effects of 
simplifications of the
hydrostatic equilibrium equation based on assuming a depth-independent
vertical gravity acceleration.

In the present paper, we will consider models of vertical structure of AGN
disks in detail. In particular, we will study the dependence of computed
vertical structure and corresponding emergent spectrum on the adopted value of
viscosity and on the degree of sophistication of the modeling procedure.
In order to better emphasize the observable
consequences of self-consistent, non-LTE models, we present the predicted
spectra for a few representative points on the disk. Complete spectra
which are obtained by integrating the local spectra over the disk surface,
taking into account general relativistic photon transfer functions
(Cunningham 1975; Speith, Riffert, \& Ruder 1995; Agol 1997; 
Agol, Hubeny, \& Blaes 1998), will
be considered in subsequent papers of this series.

Our aim here is not to construct a model to be used for comparison with
actual observations. Instead, we intend to study the sensitivity of
computed models on the degree of approximations used in the
modeling procedure. Therefore, we will first give a detailed overview of
the structural equations and the modeling procedures. This is meant to
provide a firm framework on which our approach is based, which in turn 
is useful to assessing possible systematic effects within our models.

%
%

\section{BASIC EQUATIONS AND APPROXIMATIONS}
\label{eq}

\subsection{Relativistic Disk Structure}
\label{eq_gr}

We assume a steady--state, thin, rotationally symmetric 
disk. We define the thin disk by the two conditions, namely 
$a$) the disk  height, $h(R)$, at radial distance $R$ is much smaller 
than $R$, i.e., $h(R) \ll R$, and 
$b$) all the components of the energy flux vector are
negligible to that in the vertical direction, 
i.e., the direction perpendicular 
to the disk plane. Vertical distance from the disk plane is denoted $z$.
We also assume the standard disk model, in which the azimuthal velocity is
much larger than the radial velocity, and the vertical velocity component is
neglected altogether.

The equations for relativistic disk structure were derived by Novikov 
\& Thorne (1973), Page \& Thorne (1974), Eardley \& Lightman (1975) -- 
who have corrected a previously incorrect term in the vertical pressure 
balance, and recently by Riffert \& Herold (1996) whose results we use here.

The four basic equations describing the disk structure are:

\noindent i) {\em Vertical hydrostatic equilibrium},
\begin{equation}
\label{he_gen}
{\partial P \over \partial z} = - \rho z \, {GM\over R^3}\,  {C \over B}\, ,
\end{equation}
where $P$ is the total pressure,  $\rho$ is the mass density,
$G$ is the gravitational constant, $M$ is the black hole mass, and $B$ and $C$ 
(together with $A$ and $D$ used later) are the so-called relativistic 
corrections -- see below.
We note that Abramowicz, Lanza, \& Percival (1997) have recently
rederived the vertical hydrostatic equilibrium equation, and showed that
previous treatments yielded some unphysical singularities close to the last
stable orbit.

\noindent ii) {\em Energy balance equation},
\begin{equation}
\label{ener_gen}
{\partial F_z \over \partial z} = {3 \over 2}\,  \sqrt{{GM\over R^3}}\, 
   {A \over B}\, t_{\phi r}\, ,
\end{equation}
where $F_z$ is the $z$--component of the energy flux, and $t_{\phi r}$ is 
the sheer stress (also called the viscous stress).

\noindent iii) {\em  Azimuthal momentum balance} (written using the equation
of continuity, and with the boundary condition $t_{\phi r}=0$ at the
innermost stable orbit),
\begin{equation}
\label{mom_gen}
\int_{-h}^h t_{\phi r}\, dz = {\dot M \over 2\pi}\,  \sqrt{{GM\over R^3}}\,  
{D \over A} \, ,
\end{equation}
where $\dot M$ is the mass accretion rate.

\noindent iv) Finally, the {\em equation describing the source of viscous
stress},
\begin{equation}
\label{visc_gen}
t_{\phi r} =  {3 \over 2}\, \eta \,  \sqrt{{GM\over R^3}} \, {A \over B} \, ,
\end{equation}
which specifies the viscous stress in terms of velocity gradients and 
the sheer viscosity, $\eta$.

The relativistic corrections are given by
\begin{equation}
\label{rha}
A = 1 - {2 \over r} + {a^2 \over r^2}\, ,
\end{equation}
\begin{equation}
\label{rhb}
B = 1 - {3 \over r} + {2 a \over r^{3/2}}\, ,
\end{equation}
\begin{equation}
\label{rhc}
C = 1 - {4 a \over r^{3/2}} + {3 a^2 \over r^2}\, ,
\end{equation}
\begin{equation}
\label{rhd}
D  = {1 \over 2 \sqrt r} \int_{r_i}^r {x^2 - 6x + 8a \sqrt x - 3 a^2
     \over \sqrt x \left( x^2 - 3x + 2a \sqrt x \right)}\,  dx\, ,
\end{equation}
where $r$ is radius of the annulus 
expressed in units of the gravitational radius, $R_{\rm g} = GM/c^2$,
i.e.,
\(
r = R/(GM/c^2)
\);
and the specific angular momentum $a/M$ is expressed in units of $G/c$;
$c$ being the speed of light, and
$r_i$ the radius of the innermost stable orbit.

Equations (\ref{he_gen}) - (\ref{visc_gen}) are general equations which
describe the structure of the disk. To make them applicable to real disks,
one has to specify, in addition, 
$a$) the nature of the energy flux
(i.e. how the energy generated by viscous sheer is transported), and
$b$) the nature (and numerical value) of the sheer viscosity.

The first item is straightforward. It is usually assumed that the energy is
transported solely by radiation (plus possibly in part by convection in
convectively unstable layers). In this case $F_z$ is the total radiation flux.
From Eq. (\ref{ener_gen}) it follows that
\begin{equation}
\label{flux_top}
F_z(z\! =\! h) =  {3 \over 2}\,  \sqrt{{GM\over R^3}} \, {A \over B}
  \int_0^h t_{\phi r}\, dz\, ,
\end{equation}
because the flux vanishes at the disk plane, $F_z(z\!=\!0) = 0$, since the disk 
is symmetric about the disk midplane. 
It is customary to express the total energy flux at the disk surface through
the {\em effective temperature}.  As follows from Eqs.~(\ref{flux_top})
and (\ref{mom_gen}), 
\begin{equation}
\label{teff_def}
F_z(z\! =\! h) \ \equiv\  \sigma \,  T_{\rm eff}^4\  =\ 
{3\over 8\pi} {G M \dot M \over R^3}\,  {D \over B}\, ,
\end{equation}
where $\sigma$ is the Stefan--Boltzmann constant.
The effective temperature has thus the meaning of the temperature for which
the corresponding black-body radiation has the total radiation energy
(integrated over all frequencies) equal to the total energy generated in 
the column of unit cross-section in the disk.
This explains why the early approaches used the black-body radiation for
describing the disk radiation; however, we stress that
the radiation emergent from the disk does not have to possess the
black-body frequency distribution.

%
%

\subsection{Viscosity}
\label{eq_visc}

Viscosity is the most uncertain physical quantity 
of the accretion disk modeling. There is no theory
that would explain the accretion disk viscosity form first principles,
although, at least in the case of accretion disks in the cataclysmic
variables, the Balbus--Hawley (1991) instability and subsequent detailed
numerical MHD simulations (e.g., Stone et al. 1996) represent a breakthrough 
in our understanding of disk viscosity, and offer a great promise for the
future. 

In this paper, we adopt the traditional approach, and parameterize the viscosity
by means of some adjustable parameters. First, we express the sheer
viscosity through the {\em kinematic viscosity}, $w$, as
\begin{equation}
\label{kin_visc}
\eta \equiv \rho w\, .
\end{equation}
We also introduce the {\em vertically averaged kinematic viscosity},
\begin{equation}
\label{barw_def}
\lbar w = {\int_0^h w \rho \, dz \over \int_0^h \rho \, dz} = 
               {1 \over m_0} \int_0^h \eta \, dz =
               {1 \over m_0} \int_0^{m_0} w \, dm\, ,
\end{equation}
where we have introduced the column mass, $dm =-\rho \, dz$, i.e., $m(z)$ is
the column mass above the height $z$. The total column density at the disk
midplane is denoted by $m_0$, and is related to the traditional disk surface
density, $\Sigma$, by $m_0 = \Sigma/2$.

The most commonly used viscosity parameterization is the so-called
$\alpha$--prescription (Shakura \& Sunyaev 1973). In this model, viscosity is 
thought to be caused by turbulence; the kinematic viscosity is then postulated, 
by a simple dimensional analysis, to be equal to
\begin{equation}
\label{visc_alpha1}
w = \alpha \, l_{\rm turb}\, v_{\rm turb}\, ,
\end{equation}
where $l_{\rm turb}$ and  $v_{\rm turb}$ are the size and velocity of the 
largest turbulent cells, respectively, and $\alpha$ is an ad hoc proportionality
constant. 
There are several variants of the
$\alpha$-prescription, depending of what is taken for $l_{\rm turb}$ and 
$v_{\rm turb}$. Typically, $l_{\rm turb}$ is taken to be equal to $h$, the
disk height, and $v_{\rm turb}$ equal to the sound speed, 
$c_{\rm s} = \sqrt{(P/\rho)}$, or some other turbulent velocity.
This is a {\em local} description. In fact, the original Shakura-Sunyaev
$\alpha$ was introduced to describe {\em vertically--averaged} quatities,
and we use this description here as well. We define
\begin{equation}
\label{tfr_av}
\lbar{t_{\phi r}} \equiv {1\over h} \int_0^h t_{\phi r}\, dz = 
\alpha_0 \, \lbar{P}
\, ,
\end{equation}
where $\lbar{P}$ is the vertically-averaged total pressure. Further, we
write
\begin{equation}
\label{tfr_av2}
\int_0^h t_{\phi r}\, dz = h \, \alpha_0 \,\lbar{P} = m_0\, \alpha_0 \, 
(\lbar{P}/\lbar{\rho}) \, ,
\end{equation}
where $\lbar\rho = m_0/h$ is the vertically averaged density.

In this paper, we do not consider $\lbar P$ and $\lbar\rho$ to be the 
model-dependent averages of actual pressure and density. Instead, the factor
$(\lbar{P}/\lbar{\rho})$ is taken as a known function of the basic 
parameters of the disk and the radius of the annulus, having the value
corresponding to the case of radiation-pressure-dominated disks,
$(\lbar{P}/\lbar{\rho}) = (\lbar{P}/\lbar{\rho})_{\rm rad}$. This
value is given by (see Sect 3.1 for a detailed derivation)
\begin{equation}
\label{bar_prho}
{\lbar{P} \over \lbar{\rho} } = {3 G M \dot M \over R^3} \, \left(
{\sigma_{\rm e} \over 8 \pi m_H c} \right)^2 \, {D^2 \over B C} \, ,
\end{equation}
where $\sigma_{\rm e}$ is the electron (Thomson) scattering cross-section,
and $m_H$ is the mass of hydrogen atom. The vertically averaged kinematic
viscosity then follows from integrating Eq.~(\ref{visc_gen}) over $z$, and
using Eqs.~(\ref{barw_def}), (\ref{tfr_av2}) and (\ref{bar_prho}),
\begin{equation}
\label{barw_alp}
\lbar{w}  = 2 \dot M^2 \alpha_0 \, \sqrt{{G M  \over R^3}}\left(
{\sigma_{\rm e} \over 8 \pi m_H c} \right)^2 \, {D^2 \over A C} \, .
\end{equation}
The advantage of this parameterization of viscosity is that it yields the
total column mass as an explicit function of radial distance.
Substituting Eqs.~(\ref{tfr_av2}) and (\ref{bar_prho}) into (\ref{mom_gen}),
we obtain
\begin{equation}
\label{m0_alp}
m_0  = {16 \pi \over 3} \, \left( {m_H c \over \sigma_{\rm e}} \right)^2
 \sqrt{{ R^3 \over G M }}\, {1 \over \dot M \alpha_0 }
\, {B C \over A D} \, .
\end{equation}
Because of this feature, we can easily use the mass column
density, $m$, as an independent depth variable of the problem. This has a
significant benefit of greater numerical stability of the solution of
the structural equations, in particular the radiative transfer and the
hydrostatic equilibrium equations -- see below.

Alternatively, one may parameterize the vertically--averaged kinematic 
viscosity through the Reynolds number, $Re$, which is an approach suggested 
already by Lynden-Bell \& Pringle (1974),
\begin{equation}
\label{rey0}
\lbar w = {\sqrt{G M R} \over Re }\, ,
\end{equation}
in which case
\begin{equation}
\label{totm}
m_0 = {\dot M \, Re \over 6 \pi \sqrt{G M R}} \, {B D \over A^2}\, ,
\end{equation}
which follows directly from Eqs.~(\ref{mom_gen}) and
(\ref{rey0}). The relation between $\alpha_0$ and $Re$ follows from
Eqs.~(\ref{rey0}) and (\ref{barw_alp}),
\begin{equation}
\label{re_alp}
Re = 2\,  \left( {4 \pi m_H c \over \sigma_{\rm e}} \right)^2 
\left( {R \over \dot M} \right)^2\, {1 \over \alpha_0} \, 
{A C \over D^2} \, ,
\end{equation}

The (depth-dependent) viscosity $w$ is allowed to vary as a step-wise 
power law of the mass column density, viz. 
\begin{equation}
\label{visc1}
w(m) = w_0 \left( {m/m_0} \right)^{\zeta_0}\, , \quad m>m_{\rm d} \, ,
\end{equation}
\begin{equation}
\label{visc2}
w(m) = w_1 \left( {m/m_0} \right)^{\zeta_1}\, , \quad m<m_{\rm d} \, ,
\end{equation}
where $m_{\rm d}$ is the division point.
In other words, we allow for a different power-law exponent for inner and
outer layers. This represents a generalization
of an approach we used previously, based on a single power-law representation
introduced by K\v r\' \i \v z \& Hubeny (1986). 
The reason for choosing a two-step
power law is that with a single power law we typically obtain a density
inversion in the deep layers, while the main reason for introducing the
power-law representation of viscosity in the first place was to  
prevent the ``thermal catastrophe'' 
of the disk in the low optical depth regions where the cooling due to strong 
resonance lines of light metals is important (see, e.g., Shaviv \&
Wehrse 1986; Hubeny 1990a).  On the other hand, recent numerical simulations
(Stone et al. 1996) indicate that that the viscosity actually increases
towards the surface, giving rise to a high-temperature region (corona)
on the top of a disk.  Such models were considered for instance by 
Sincell \& Krolik (1997).  We plan to study such non-standard 
models in future papers of this series; in this paper we will consider models
where the viscosity decreases in the outer layers.

We have thus four independent parameters:
exponents $\zeta_0$ and $\zeta_1$, the division point, $m_{\rm d}$, and the
fraction, $f$, of energy dissipated in deep layers, $m>m_{\rm d}$. 
The coefficients
$w_0$ and $w_1$ are derived from the condition on the vertically averaged
viscosity, 
$ \int_0^{m_0} w(m) dm/m_0 = \lbar w $, and
$ \int_{m_{\rm d}}^{m_0} w(m) dm/m_0 = f \, \lbar w $. We obtain
\begin{equation}
\label{w0}
w_0 = {f \, \lbar w (\zeta_0 + 1) \over 1 -(m_{\rm d}/{m_0})^{\zeta_0 + 1} }\, ,
\end{equation}
\begin{equation}
\label{w1}
w_1 = {(1-f)\, \lbar w (\zeta_1 + 1) \over 
(m_{\rm d}/{m_0})^{\zeta_1 + 1}}\, .
\end{equation}
Generally, $w(m)$ does not have to be continuous at the division point
$m_{\rm d}$. If we require the continuity, then $f$ and $m_{\rm d}$ are no 
longer two independent parameters; instead, they are related through
\begin{equation}
\label{mdiv}
{m_{\rm d}\over m_0} = \left( 1 + {\zeta_0 + 1 \over \zeta_1 + 1 } 
{f \over 1 - f} \right)^{-{1 \over \zeta_0 + 1}} \, .
\end{equation}
Typically, the deep-layer power law exponent $\zeta_0$ is set to 0 
(constant viscosity), while the ``surface'' power law exponent $\zeta_1$ is 
usually set to a value larger than zero. 
In \S 3.3, we will compare this treatment
to a purely local $\alpha$--parameterization of viscosity.

%
%

\subsection{Equations for the Vertical Structure}
\label{eq_vert}

Due to the assumption of thin disk, we may reduce a general 2-D problem
of computing disk structure to a set of 1-D problems.
The disk is divided into a set of axially symmetric concentric annuli; 
each annulus behaves as a one-dimensional radiating slab.
The vertical structure of a single annulus is computed by solving simultaneously
the hydrostatic equilibrium equation, the energy balance equation, the
radiative transfer equation, and, since we do not generally assume LTE, 
the set of statistical equilibrium equations. Below, we specify these
equations in detail. Since the state parameters now depend only on the 
vertical distance $z$, we replace partial derivatives in Eqs.~(\ref{he_gen})
and (\ref{ener_gen}) by ordinary derivatives. 
Moreover, we take the column density $m$ as the basic depth variable, as it
is customary in modeling stellar atmospheres and non-relativistic accretion
disks.
Furthermore, we write down the terms corresponding to radiation
(i.e., radiation pressure, and radiation flux) explicitly, to stress that we
treat the radiation essentially exactly, without any simplifying assumptions
about radiative transfer. Since the radiation terms in the structural
equations are written using the radiative transfer equation, we start with
it.

\noindent{\em a) Radiative Transfer Equation}
 
\noindent The radiative transfer equation is written in the standard way 
(e.g. Mihalas 1978), viz
\begin{equation}
\label{rte}
\mu \, {d I_\nu(\mu) \over d\tau_\nu} = I_\nu - S_\nu  \, ,
\quad {\rm or} \quad 
\mu \, {d I_\nu(\mu) \over dm} = -{1 \over \rho} \, \left(\chi_\nu I_\nu - 
\eta_\nu \right) \, , 
\end{equation}
where $I_\nu(\mu)$ is the specific intensity of radiation at frequency
$\nu$, 
$\mu$ being the cosine of the angle between direction of propagation and
the normal to the disk midplane.
The monochromatic optical depth is defined through 
$d \tau_\nu \equiv -\chi_\nu dz = (\chi_\nu/\rho) dm $; 
and the source function by $S_\nu \equiv \eta_\nu/\chi_\nu$.
Here, $\chi_\nu$ is the total
absorption coefficient, $\chi_\nu= \kappa_\nu+\sigma_\nu$, $\sigma_\nu$
being the scattering coefficient. We assume that the only
scattering process is the electron (Thomson) scattering, 
$\sigma_\nu = \sigma_{\rm e}$; $\sigma_{\rm e}$ being the Thomson cross-section. 
Finally, $\eta_\nu$
is the emission coefficient, given by
$\eta_\nu = \eta_\nu^{\rm th} + \sigma_\nu J_\nu$,
where $\eta_\nu^{\rm th}$ is the coefficient of thermal emission.
We assume the symmetry condition
at the disk midplane, $I_\nu(\mu)=I_\nu(-\mu)$.
The upper boundary condition is $I_\nu(\mu)=I^{-}_\nu, \mu <0$, where
$I^{-}$ is a prescribed incident radiation. In the present paper we assume,
for simplicity, the case of no external irradiation, $I^{-}=0$. We return to 
the problem of non-zero external irradiation in a future paper of this series.

The first two moment equations of the transfer equation read
\begin{equation}
\label{rtem1}
{d H_\nu \over dm} = - {1 \over \rho}\, \left( \kappa_\nu J_\nu - \eta_\nu
\right) \, ,
\end{equation}
\begin{equation}
\label{rtem2}
{d K_\nu \over dm} = - {\chi_\nu \over \rho}\, H_\nu  \, ,
\end{equation}
where the moments of the specific intensity are defined by
\begin{equation}
\label{kdef}
[J_\nu,\,  H_\nu, \, K_\nu ] = {1 \over 2} \int_{-1}^1 [1,\, \mu,\, \mu^2] \, 
I_\nu(\mu)\, d\mu \, ,
\end{equation}
We stress that in Eq.~(\ref{rtem1}) the scattering terms cancel, so that
the equation contains the true absorption coefficient $\kappa_\nu$, instead of
the total absorption coefficient $\chi_\nu$.
The radiation flux is given by 
\begin{equation}
\label{rflux}
F_{\rm rad} = 4 \pi \int_0^\infty \! H_\nu d\nu \, . 
\end{equation}

To solve the radiative transfer equation, we employ the Variable
Eddington Factors technique (Auer \& Mihalas 1970), which consists in
introducing the Eddington factor $f_\nu^K = K_\nu/J_\nu$, and writing
the transfer equation (\ref{rte}) as
\begin{equation}
\label{rtevef}
{d^2 \!\left(f_\nu^K J_\nu \right) \over d\tau_\nu^2} = J_\nu - S_\nu  \, .
\end{equation}
This equation involves only the mean intensity of 
radiation, $J_\nu$, which is a function of only frequency and depth,
and not the specific intensity which in addition is a function of
angle $\mu$. Such an approach is extremely advantageous in 
methods which solve the global system of structural equations
iteratively.
The Eddington factor is determined by a set of
frequency-by-frequency formal solutions of the transfer equation, and
is held fixed at a subsequent iteration step of the global scheme.

\noindent{\em b) Hydrostatic Equilibrium}

\noindent The hydrostatic equilibrium equation reads, neglecting 
self-gravity of
the disk and assuming that the radial distance from the black hole, $R$,
is much larger than the vertical distance from the central plane, $z$, 
\begin{equation}
\label{he}
{dP \over dm} =  g(z)\, ,
\end{equation}
where the depth-dependent vertical gravity acceleration is given by
\begin{equation}
\label{gdef}
 g(z) = {GM \over R^3 }{C \over B} \, z \, .
\end{equation}
The total pressure is given as a sum of the gas pressure and the radiation 
pressure,
\begin{equation}
\label{press}
P=P_{\rm gas} + P_{\rm rad} = NkT + 
{4\pi\over c} \int_0^\infty K_\nu \, d\nu \, ,
\end{equation}
where $N$ is the total particle number density, $T$ the temperature,
$k$ the Boltzmann constant. 
The upper boundary condition is taken from Hubeny (1990a -- Eqs. 4.19--4.20 
there).

\noindent{\em c) Energy Balance}

\noindent Substituting Eqs. (\ref{visc_gen}) and (\ref{kin_visc}) into 
(\ref{ener_gen}), and using Eqs. (\ref{rflux}) and (\ref{rtem1}),
we obtain
\begin{equation}
\label{ener}
{9 \over 4} {G M\over R^3} \left( {A \over B} \right)^2 \, \rho \,w
=4 \pi \int_0^\infty (\eta_\nu - \kappa_\nu J_\nu) d\nu  \, .
\end{equation}
The energy balance equation may be cast to different form if we do not express
the radiation flux through the moment equation of the transfer equation,
namely
\begin{equation}
\label{ener2}
{d F_{\rm rad} \over dm } = - {9 \over 4} {G M\over R^3} 
\left( {A \over B} \right)^2 \, w(m) \, .
\end{equation}
Since the radiation flux at the midplane, $m = m_0$, is zero, equation
(\ref{ener2}) may be integrated to yield
\begin{equation}
\label{ener2a}
F_{\rm rad}(m)  =  {9 \over 4} {G M\over R^3} 
\left( {A \over B} \right)^2 \, \int_m^{m_0} w(m') \, dm' \, .
\end{equation}
Using Eqs. (\ref{barw_def}) -- (\ref{totm}), and the definition of effective
temperature, Eq. (\ref{teff_def}), we obtain finally
\begin{equation}
\label{ener3}
F_{\rm rad}(m)  =  \sigma T_{\rm eff}^4 \, [1 - \theta(m)] \, ,
\end{equation}
where the auxiliary function $\theta$ is defined by
\begin{equation}
\label{theta}
\theta(m) = {1 \over \lbar w \, m_0}\, \int_0^{m} w(m') \, dm' \, .
\end{equation}
For any depth dependence of viscosity, $\theta$ is a monotonically
increasing function of $m$, between $\theta(0) = 0$, and $\theta(m_0) = 1$.
For a depth-independent viscosity, $\theta(m)= m/m_0$, and for the
adopted step-wise power law defined by Eqs. (\ref{visc1}) -- (\ref{mdiv}),
it is given by
\begin{equation}
\label{theta1}
\theta(m) = (1-f)\, (m/m_{\rm d})^{\zeta_1 + 1}\, , \quad m\leq m_{\rm d} \, , 
\end{equation}
\begin{equation}
\label{theta2}
\theta(m) = (1-f) + f \, 
 {(m/{m_0})^{\zeta_0 +1}-(m_{\rm d}/{m_0})^{\zeta_0 +1}
\over 1 - (m_{\rm d}/{m_0})^{\zeta_0 + 1}}\, ,  \quad m\geq m_{\rm d} \, .
\end{equation}

\noindent{\em d) The z--m relation}

\noindent Since the hydrostatic equilibrium equation (\ref{he}) contains the 
vertical distance $z$ explicitly, we have to supply the relation between 
$z$ and $m$, which reads simply
\begin{equation}
\label{z_m}
{dz \over dm} = -{1 \over \rho}\, .
\end{equation}

\noindent{\em e) Absorption and emission coefficient}

\noindent The absorption coefficient is given by
\begin{eqnarray}
\label{abso}
\chi_\nu & = & \sum_i \sum_{j>i} \left[ n_i - (g_i/g_j) n_j \right]
\sigma_{ij}(\nu)
+ \sum_i \left( n_i - n_i^\ast e^{-h\nu/kT} \right) \sigma_{i\kappa}(\nu) 
\nonumber \\
 & & + \sum_\kappa n_{\rm e} n_{\kappa} \sigma_{\kappa\kappa}(\nu,T)
\left( 1 - e^{-h\nu/kT} \right) + n_{\rm e} \sigma_{\rm e} \, ,
\end{eqnarray}
where $n_i$ is the number density (population) of an energy level $i$
(we number all levels consecutively, without notational distinction of
the parent atom and ion), $n_i^\ast$ the corresponding LTE population,
$g_i$ the statistical weight, and $\sigma(\nu)$ the appropriate
cross-section.
The four terms represent, respectively, the contributions of bound-bound
transitions (i.e. spectral lines), bound-free transitions (continua), free-free
absorption (inverse brehmstrahlung), and of electron scattering.
Subscript $\kappa$ denotes the ``continuum'', and $n_\kappa$ the ion number
density. The negative contributions in the first three terms represent the
stimulated emission. There is no stimulated emission correction for the
scattering term, since this contribution cancels with ordinary 
absorption for coherent scattering (for an illuminating discussion, 
see Shu 1991).

Analogously, the thermal emission coefficient is given by
\begin{eqnarray}
\label{emis}
\eta_\nu^{\rm th} & = & \left( 2h\nu^3/c^2 \right) \Bigl[ \,
\sum_i \sum_{j>i}  n_j (g_i/g_j) \sigma_{ij}(\nu)
+ \sum_i n_i^\ast  \sigma_{i\kappa}(\nu) \, e^{-h\nu/kT} \nonumber \\
 & & + \sum_\kappa n_{\rm e} n_{\kappa} \sigma_{\kappa\kappa}(\nu,T) \,
 e^{-h\nu/kT} \, \Bigr] \, .
\end{eqnarray}
The three terms again describe the bound-bound, bound-free, and free-free
emission processes, respectively.

These equations should be complemented by expressions for the relevant
cross-sections, definition of LTE populations, and other necessary expressions.

We did not yet implement the Compton scattering in our codes, but
the work on this problem is under way, and will be
reported in a future paper. 
Nevertheless, Compton scattering is not important for the models considered 
in this paper, as we shall verify in Sect. 3.1.

\noindent{\em f) Statistical Equilibrium Equation}

\noindent It is well known that the LTE approximation breaks down in 
low-density, radiation-dominated media (see, e.g. Mihalas 1978), which are 
precisely the  conditions prevailing in the AGN disks. Therefore, we have to 
adopt a more general treatment, traditionally called non-LTE (or NLTE), where 
the populations
of some selected energy levels of some selected atoms/ions are allowed to
depart from the Boltzmann-Saha distribution. These populations are determined
through the equations of statistical equilibrium (e.g. Mihalas 1978), viz.
\begin{equation}
\label{ese1}
n_i \sum_{j \ne i} \left( R_{ij} + C_{ij} \right) = 
\sum_{j \ne i} n_j\left( R_{ji} + C_{ji} \right) \, ,
\end{equation}
where $R_{ij}$ and $C_{ij}$ is the radiative and collisional rates,
respectively,
for the transition from level $i$ to level $j$. The l.h.s. of (\ref{ese1})
represents the total number of transitions out of level $i$, 
while the r.h.s. represents the total number of transitions into 
level $i$ from all other levels.
An essential numerical complication inherent to this approach is that the
radiative rates are given as integrals over the radiation intensity, i.e.,
schematically
\begin{equation}
\label{rij}
R_{ij} = \int_0^\infty (h\nu/4\pi) \, \sigma_{ij}(\nu)\, J_\nu \,d\nu\, .
\end{equation}

\subsection{Numerical Method}
\label{num}

The overall system of Eqs. (\ref{rtevef}), (\ref{he}), (\ref{ener}), 
(\ref{z_m}), and (\ref{ese1}), together with auxiliary expressions 
(\ref{gdef}), (\ref{rey0}) -- (\ref{mdiv}),
and the definition expressions for the absorption 
and emission coefficients, (\ref{abso}) and (\ref{emis}),
form a highly coupled, non-linear set of integro-differential equations. 
Fortunately, these equations are very similar
to the equations describing classical NLTE stellar atmospheres
(see, e.g., Hubeny 1990a,b), where the modeling techniques are
highly advanced. We may therefore employ to great advantage numerical methods
and computer programs designed originally for stellar atmospheres.

We use here the computer program TLUSDISK, which is a derivative of the stellar
atmosphere program TLUSTY (Hubeny 1988). The program 
is based on the
hybrid complete-linearization/accelerated lambda iteration (CL/ALI) method
(Hubeny \& Lanz 1995). The method resembles the traditional 
complete linearization,
however the radiation intensity in most (but not necessarily all) frequencies
is not linearized; instead it is treated via the ALI scheme (for a review
of the ALI method, see e.g., Hubeny 1992).
A NLTE model of a vertical structure of one annulus of a disk is 
generally computed in several
steps. First, an LTE-gray model is constructed, as described in Hubeny (1990a).
This serves as the starting solution for the subsequent step, 
an LTE model, computed by TLUSDISK. This in turn is used as the starting
solution for the next step, a NLTE model.

In the NLTE step, we first calculate the so-called NLTE/C model
(i.e., NLTE with continua only), assuming 
that all bound-bound transitions are in detailed radiative balance. 
Finally, in the last step, we consider all lines -- we denote this model as
NLTE/L. The lines influence the disk structure both directly -- 
through their contribution to the total heating/cooling rate 
and to the total radiation pressure, as well as indirectly --
through their influence on atomic level population via equations of
statistical equilibrium. Therefore, they influence the heating/cooling rates
and the radiation pressure in the continuum processes as well.
However, one should be aware that considering lines is not, strictly 
speaking, consistent with the 1-D approach adopted here.
The assumption of horizontally homogeneous rings implies
that the Keplerian velocity is assumed constant within the ring.
This is a good approximation for continua, but
may be inaccurate  for the radiation transfer in spectral lines because a 
difference of projected Keplerian velocity as small as 
the thermal velocity already shifts the line photon out of the Doppler core. 
The photon escape probability from a disk ring may therefore be quite different
compared to the escape probability from a plane-parallel, static atmosphere.
For a proper treatment of this problems we would have to abandon a
1-D modeling and construct a fully 2-D model. However, at present, we intend
to explore a magnitude of various phenomena and their effect on
the disk structure; we feel that for such an exploratory model study the
1-D treatment as described above is satisfactory.

%
%

\section{SENSITIVITY ANALYSIS}

As a representative case, we take a disk around a Kerr supermassive black hole
with $M = 2 \times 10^9 M_{\odot}$, with the maximum stable rotation
(the specific angular momentum $a/M=0.998$).
The mass flux is taken $\dot{M}=1~ M_\odot$/year.
We have calculated a number of vertical structure models at various radii;
we present here three representative models for
$r = 2$, 11, and 20. The corresponding effective temperature is (roughly)
80,000 K, 27,000 K, and 18,000 K, respectively; the models thus cover a 
representative range of effective temperatures of the AGN disk annuli.

For simplicity, we consider disks composed of hydrogen and helium only.
This allows us to study various effects without spending too much
computer time, while taking into account all the essential physics.
The effects of heavy elements on disk models will be considered in the
subsequent paper. 
Here, we intend to investigate two questions, namely, i) the effect
of degree of sophistication in the disk modeling (LTE versus NLTE; effects
of lines); and ii) the sensitivity of computed model structure on the 
individual viscosity parameters.

Hydrogen is represented essentially exactly: The first 8 levels are
treated separately, while the upper levels are merged into the averaged
non-LTE level accounting for level dissolution as described by Hubeny,
Hummer, \& Lanz (1994). Neutral helium is represented
by a 14-level model atom, which incorporates all singlet and triplet
levels up to $n=8$. The 5 lowest levels are included individually;
singlet and triplet levels are grouped separately from $n=3$ to $n=5$,
and we have formed three superlevels for $n=6, 7,$ and~8.
The first 14 levels of He$^+$ are explicitly treated. We assume a solar
helium abundance, $N({\rm He})/N({\rm H}) = 0.1$.

In the NLTE/L models, all the lines are treated explicitly, assuming 
Doppler profiles with turbulent velocity 
ranging from the thermal velocity up to the vertically-averaged sound speed,
\begin{equation}
v_{\rm turb} = \sqrt{\lbar{P}/\lbar{\rho}}\, ,
\end{equation}
where $(\lbar{P}/\lbar{\rho})$ is given by Eq.~(\ref{bar_prho}).
For the three annuli at $r$ = 2, 11, and 20, of the disk specified above, 
the sound speed has values of 3880, 770, and 375 km s$^{-1}$, respectively. 
Although considering different values of $v_{\rm turb}$ has a significant 
effect on computed rest-frame line profiles, we found a negligible effect 
on the resulting disk structure.

\subsection{Non-LTE Effects}
\label{lte-nlte}

Figure 1 shows the temperature as a function of depth for the three annuli.
The depth is expressed as column mass in g cm$^{-2}$.
We see several interesting features. 
Firstly, for the hot model at $r=2$ ($T_{\rm eff} =80,000$ K), 
the NLTE temperature
structure differs appreciably from the LTE structure, even at the
midplane. The cooler models exhibit different temperature at the
upper layers, i.e. for $\log m \lta 1.5$, while the temperature
structure is unchanged by NLTE effects in the deep layers. This is explained 
by the fact that the effective optical thickness 
\[ 
\tau_{\rm eff} \approx \sqrt{\tau_{\rm tot}\, \tau_{\rm abs}} \, ,
\]
(where $\tau_{\rm tot}$ is the total optical depth corresponding to $\chi$, 
i.e., including electron scattering, while $\tau_{\rm abs}$ is the optical 
depth corresponding to the true absorption, $\kappa$), becomes comparable to or
smaller than 1 for hot models. In other words, a photon created at the 
midplane has a non-zero probability that it will undergo a series of 
consecutive scatterings without being destroyed by a thermal process until 
it escapes from the disk surface.
Consequently, even the deep layers of the disk
now effectively ``feel'' the presence of the boundary, which gives rise to
NLTE effects even close to the midplane.

Secondly, as expected, differences between NLTE/C and NLTE/L models are 
negligible in the deep layers, while they are important in the upper layers --
the lines heat up the upper disk atmosphere
(for $\log m \lta 0.5$) by some 10,000 K for the hot model; by 4,000 K
for the intermediate ($r=11$) model, and by only 1000 K for the ``cool''
model. The effect is exactly analogous to the temperature rise predicted in hot
stellar atmospheres (Mihalas 1978), which is explained as an indirect
effect of Lyman and Balmer lines on the heating in the Lyman and Balmer
continuum. 
The explanation goes as follows: roughly speaking, the temperature structure
is determined by the balance between the radiative heating, which is mostly 
provided by the Lyman (for hotter models) and the Balmer (for cooler models) 
continuum, and the radiation cooling in the free-free transitions in the 
optical and infrared region.
Radiative transfer in the Lyman (Balmer) lines explicitly gives rise 
to an overpopulation of the hydrogen $n=1$ ($n=2$) states, which leads to
an increase of the efficiency of the Lyman (Balmer) continua 
which in turn leads to an additional heating at the surface.
There is a competition between this heating and the traditional surface cooling
caused by the lines, but in the present case the indirect effect
dominates. 

In Fig. 2, we plot the density structure for the same models as displayed in
Fig.~1. For the hot model, density is almost unaffected by NLTE effects.
This is easy to understand, since in this case the radiation pressure is
completely dominant, and the disk structure is given by the following
simple analytic formulae.
The hydrostatic equilibrium equation (\ref{he}) reduces to
\begin{equation}
\label{dprad}
{d P_{\rm rad} \over d m} = Q\, z \, ,
\end{equation}
where we have denoted $Q = (G M/R^3) \, (C/B)$. Using Eq.~(\ref{press}) and the
first moment of the transfer equation, (\ref{rtem2}), we may express the
gradient of the radiation pressure as
\begin{equation}
\label{dprad_h}
{d P_{\rm rad} \over d m} = {4\pi \over c} \int_0^\infty {d K_\nu \over dm} \,
d\nu =
{4\pi \over c} \int_0^\infty {\chi_\nu \over \rho}\, H_\nu \, d\nu= 
{4\pi \over c} \chi_{H} \, H  \, ,
\end{equation}
where  $\chi_{H}$ is the flux-mean opacity, defined by
\begin{equation}
\label{chi_h}
\chi_{H} = \int_0^\infty (\chi_\nu/\rho) H_\nu \, d\nu/H \, ,
\end{equation}
and $H$ is the total
(frequency-integrated) Eddington flux, 
$H = \int_0^\infty H_\nu\, d\nu = 4 \pi F_{\rm rad}$.
The flux-mean opacity is particularly simple in the case when the
electron scattering is the dominant source of opacity. In this case
$\chi_H = n_{\rm e} \sigma_{\rm e}/\rho$. For a medium where hydrogen
is fully ionized and helium is partially doubly
ionized, the flux-mean opacity is given by 
$\chi_H = (\sigma_{\rm e}/m_H)(1+\alpha Y)/(1 +4 Y) 
\approx 0.4\, (1+\alpha Y)/(1 +4 Y)$, where $Y$ is the
helium abundance, $Y= N({\rm He})/N({\rm H})$, and 
$\alpha = 1 + N({\rm He}^{++})/N({\rm He})$; $m_H$ is the mass of hydrogen
atom.

Using Eq.~(\ref{ener3}), we obtain
\begin{equation}
\label{enerz}
{\sigma \chi_H \over c}\, T_{\rm eff}^4 \, [1 - \theta(m)] \, = \, Q \, z \, .
\end{equation}
We know that at the disk surface, $m=0$, $\theta(m)=0$, and therefore the
total disk height, $h$, is given by
\begin{equation}
\label{totz}
h = {\sigma T_{\rm eff}^4 \over Q} {\chi_H \over c}\  =\ 
{3 \dot M \chi_H \over 8 \pi c} {D \over C} \, .
\end{equation}
The hydrostatic equilibrium equation thus reduces to 
\begin{equation}
\label{z-theta}
{z \over h} \ = \ 1 - \theta(m)\, 
\end{equation}
This equation enables us to estimate the density, since from Eq.~(\ref{z_m})
we know that $1/\rho = - dz/dm$, so that from Eq.~(\ref{z-theta}) we
obtain $1/\rho = h\, (d\theta/dm)$. However, from the definition of $\theta$,
Eq.~(\ref{theta}) we have $d\theta/dm = w(m)/(\lbar w m_0)$, so that we
finally obtain
\begin{equation}
\label{rho-m}
\rho(m)\ =\ {m_0\over h}\, {\lbar w\over w(m)} \, .
\end{equation}
This equation shows that the density structure for the case of dominant
radiation pressure is solely determined by the dependence of viscosity
on depth. For a depth-independent viscosity, density is constant, and
is given by $\rho = m_0/h$, i.e the total column mass divided by the disk
height. 

Before proceeding further, we derive the term $(\lbar{P}/\lbar{\rho})$ for the
case of radiation-pressure dominated disks. Assuming a depth-independent
viscosity, the density is also constant, $\rho = \lbar{\rho}$. The 
hydrostatic equilibrium equation thus reads $dP/dz = -Q\, \lbar{\rho}\, z$,
which has solution $P(z) = Q\, \lbar{\rho}\, (h^2 - z^2)/2$, and
consequently $\lbar{P} = Q\, \lbar{\rho}\, h^2/3$. Substituting for $h$
from Eq.~(\ref{totz}), we finally obtain
\begin{equation}
\label{bar_prho0}
{\lbar{P}\over \lbar{\rho}} =  {3 G M \dot M \over R^3} \, \left( {\chi_H
 \over 8 \pi m_H c} \right)^2 \, {D^2 \over B C} \, .
\end{equation}
Substituting the pure-hydrogen form of $\chi_H$, $\chi_H= \sigma_{\rm e}/m_H$
into Eq.~(\ref{bar_prho0}), we obtain Eq.~({\ref{bar_prho}) used in the 
definition of $\alpha_0$.

It should be stressed that Eqs.~(\ref{dprad}) through (\ref{rho-m}) apply
if the {\em gradient} of the radiation pressure
dominates over the gradient of the gas pressure,
$d P_{\rm rad}/dm \gg  d P_{\rm gas}/dm $; it is not sufficient that
$P_{\rm rad} \gg  P_{\rm gas}$.
It is clear that even in the case where
the radiation pressure dominates everywhere in the atmosphere, the radiation
pressure gradient becomes very small in the upper layers where the
optical depth is smaller than unity, because the radiation field is already
formed and does not change when going outward. The gas pressure gradient
then takes over in these layers, and consequently density starts  to decrease
exponentially with height. For a comprehensive discussion, see Hubeny (1990a).

In any case, Eq.~(\ref{rho-m}) shows that in the inner layers of the
radiation--pressure--dominated disks, density structure does depend only on 
viscosity, and is therefore insensitive to NLTE effects. This is indeed
demonstrated in Fig.~2. 
The above considerations also explain that for the 
radiation--pressure--dominated disks the disk height depends only on the
radial distance $R$, through
the relativistic correction $D/C$, and slightly through $\chi_H$ (because
of variations in the ionization of helium). The
$z$-$m$ relation is given through the depth dependence of viscosity, 
and is thus insensitive to local temperature, as well as to NLTE effects.
This is demonstrated in Fig.~3, where we plot the vertical distance $z$
as a function of $m$ for the same models as displayed in Fig.~1.

Figure 4a displays the departure coefficient ($b$-factor) for the ground
state of hydrogen for the three annuli, and for NLTE/C and NLTE/L models.
Departure coefficient is defined as $b_i = n_i/n_i^\ast$. Near the
midplane of the disk, the $b$-factor is close to unity because the optical
depth is large. It starts to deviate from unity as soon as the Lyman
continuum becomes effectively optically thin. The behavior of the hot
model differs from the two cooler ones.

In the hot model, electron scattering dominates the opacity even in the Lyman
continuum; the thermal coupling parameter 
\begin{equation}
\label{epsnu}
\epsilon_\nu = {\kappa_\nu \over \kappa_\nu +\sigma_\nu}\, ,
\end{equation}
is very small in the inner layers (e.g., $\epsilon_\nu \approx 2 \times 10^{-3}$
at the  midplane). Consequently, $J_\nu$ starts to deviate from the thermal
source function, which is roughly equal to the Planck function, already
in deep layers -- see the upper left panel of Fig.~7.
One can make these considerations more quantitative by invoking a simple
model of radiative transfer in the presence of scattering (see, e.g.,
Mihalas 1978). In a simple case of depth-independent $B_\nu$
and $\epsilon_\nu$, the mean intensity is given by
\begin{equation}
\label{japprox1}
J_\nu(\tau_\nu) \approx B_\nu(\tau_\nu) \left[ 1 +\sqrt{\epsilon_\nu} 
- \exp(-\sqrt{3\epsilon_\nu}\, \tau_\nu) \right] /(1 + \sqrt{\epsilon_\nu})\, .
\end{equation}
It is clear that $J_\nu \approx B_\nu$ for 
$\tau_\nu \gta \sqrt{3\epsilon_\nu}$ (called the thermalization depth);
while for $\tau_\nu < \sqrt{3\epsilon_\nu}$ the mean intensity 
$J_\nu$ drops below $B_\nu$; the drop is larger for smaller $\epsilon_\nu$.
We note that 
at the surface, $J_\nu(0) \approx \sqrt{\epsilon_\nu} \, B_\nu$. 

The photoionization rate is
proportional to an integral of $J_\nu$, while the recombination rate is
proportional to an integral of $B_\nu$. As a result, the recombinations
dominate over ionizations, which leads to an overpopulation of the 
hydrogen ground state. 
For cooler models, the formation of Lyman continuum is different.
The electron scattering is not overwhelmingly dominant, so the thermal 
coupling parameter $\epsilon$ is now smaller ($\epsilon$ is between $10^{-2}$
and $10^{-1}$ in the midplane layers), and the mean intensity decouples
from $B_\nu$ farther away from the midplane. The Planck function at the 
frequencies of the Lyman continuum decreases very fast with decreasing 
temperature, because the Lyman continuum frequencies are in the Wien tail of 
the Planck function, $B_\nu (T) \propto \exp (-h\nu/kT)$. 
This decrease in the local
value of $B_\nu$ is now faster than the decrease of $J_\nu$ which,
when going outward, is more and more decoupled from the local temperature.
We thus obtain $J_\nu > B_\nu$ throughout most of the atmosphere;
consequently the ionizations dominate over recombinations in the Lyman 
continuum, and hence the hydrogen ground state becomes underpopulated.

Once the Lyman continuum becomes
transparent (which happens around $\log m \approx 0$ for all models), 
$b_1$ does
not change much for NLTE/C models. For NLTE/L models, however, $b_1$
starts to increase outward above $\log m \approx 0$, which is caused by
radiative transfer in L$\alpha$ and other Lyman lines. 
This is a standard NLTE effect -- strong resonance lines 
cause the lower level to be overpopulated, while the upper level becomes
underpopulated with respect to LTE (e.g., Mihalas 1978). This is illustrated
in Fig.~4b, which displays the $b$-factor for the $n=2$ state of hydrogen. 
Detailed explanation of the behavior
of $b_2$ is quite analogous to that of $b_1$ discussed above.

The most interesting quantity predicted by the models is the emergent
radiation. Figure~5 compares the emergent flux in the Lyman limit region 
for the three models, LTE, NLTE/C, and NLTE/L. Only hydrogen and helium
lines are taken into account. In the NLTE/C model, the line source functions
are computed using NLTE level populations, which in turn were calculated by 
TLUSDISK by
considering the lines to be in detailed radiative balance. Consequently,
the predicted line profiles are inconsistent, and we show them only for
demonstration purposes.
The spectra were computed using program SYNSPEC (Hubeny, Lanz, \& Jeffery
1994), which calculates synthetic spectra for model atmospheres or disks
previously computed by TLUSTY or TLUSDISK.
We assume the turbulent velocity $v_{\rm turb}$ = 3880, 770, and 
375 km s$^{-1}$, for the annuli at $r$ = 2, 11, and 20, respectively. For
a clearer display, the final spectra are convolved with a gaussian with
FWHM = 10 \AA. We stress that the spectra are computed in the rest frame
of the annulus; to obtain spectrum received by a distant observer one
would have to take into account Doppler velocities, and general
relativistic effects (frequency shift, light bending, etc.). 
As pointed out in the Introduction, 
we will study here only the rest-frame radiation.

For the hot model, the NLTE/C and NLTE/L models give essentially the
same flux across the Lyman discontinuity which in this case virtually
disappears. In contrast, the LTE model predicts the Lyman jump
in emission. In the intermediate model, LTE predicts almost non-existent
Lyman jump while NLTE/L predicts a weak emission jump. The NLTE/C model 
predicts a larger emission jump, and spuriously strong emission in the Lyman
lines. In the cool model, LTE predicts the Lyman jump in very strong 
absorption, while the NLTE/L models again predicts essentially smooth
spectrum in this region.

The explanation of this behavior follows from the above discussion of
the hydrogen departure coefficients $b_1$ and $b_2$ (Fig.~4), 
and from studying the
behavior of $B_\nu, S_\nu, J_\nu$, and $\epsilon_\nu$ as functions of depth.
This is displayed in Fig.~7 for the hot model at $r=2$, and in Fig.~8 for
the cool model at $r=20$.
Let us first consider LTE models. In the hot model, electron scattering
completely dominates, so the total opacities at both sides of the
Lyman discontinuity are roughly equal; hence both sides of the discontinuity
are formed at the same depth. 
The only difference is the value of the thermal coupling parameter
$\epsilon_\nu$. Since the thermal opacity on the blue side ($\nu_b$) 
of the discontinuity is larger that that on the red side 
($\nu_r$), we have 
$\epsilon_{\nu_b} > \epsilon_{\nu_r}$. Consequently, the mean intensity
at the red side is more uncoupled from $B_\nu$, and consequently
$J_{\nu_b} > J_{\nu_r}$ -- see Eq.~(\ref{japprox1}). 
In fact, this is an interesting manifestation
of the classical Schuster mechanism. By the same token, one can understand
why the Lyman lines also appear in emission in the hot LTE model.

In the cool model, in contrast, we have a classical LTE jump: the 
high--opacity (blue) side of the jump is formed much higher than the
low--opacity (red) side, because the electron scattering opacity no longer
dominates over the thermal opacity. Since the 
temperature decreases outward, the blue side of the jump is formed at 
lower temperature, and the flux is consequently lower. In the intermediate 
model, both effects compete, but the first one wins, so we
obtain a weak emission jump.

NLTE effects modify this picture significantly. In the hot model,
the red side of the Lyman jump remains almost unchanged by NLTE effects
because the departure coefficient $b_2$ is close to unity around the
thermalization depth of the Balmer continuum. In the Lyman continuum,
the thermal coupling parameter $\epsilon_{\nu_b} > \epsilon_{\nu_r}$
as discussed above. However, in NLTE one has to modify Eq.~(\ref{japprox1})
to read
\begin{equation}
\label{japprox2}
J_\nu(\tau_\nu) \approx S^{\rm th}_\nu(\tau_\nu) \left[ 1 +\sqrt{\epsilon_\nu} 
- \exp(-\sqrt{3\epsilon_\nu}\, \tau_\nu) \right]/(1 + \sqrt{\epsilon_\nu})\, ,
\end{equation}
where $S^{\rm th}_\nu$ is the thermal source function. In the case of
Lyman continuum, it is roughly given by $S^{\rm th}_\nu \approx B_\nu/b_1$.
Since $b_1 > 1$ in the continuum--forming layers (see Fig.~4a), we have
$S^{\rm th}_\nu < B_\nu$. We see that the effect of an increased thermal 
coupling parameter $\epsilon_\nu$ at the Lyman continuum frequencies 
(which tends to increase the emergent intensity) is offset by a decrease of the
thermal source function. The latter results from a predominance of 
recombinations over ionizations, as explained above.

Finally, we explain why we obtain a somewhat higher flux in the Lyman
continuum for the NLTE/C models than for NLTE/L models. This is given
by the fact that $b_1({\rm NLTE/C}) < b_1({\rm NLTE/L})$ all the way
from the surface to deep layers (see Fig.~4a), which in turn is the
effect of L$\alpha$ and other Lyman lines. Because of low density, the
photon destruction parameter $\epsilon$ is rather small and therefore the
thermalization depth is rather large -- much larger than the depth of
formation (see, e.g., Mihalas 1978). Consequently, the thermal source
function in the Lyman continuum, $S^{\rm th}_\nu \approx B_\nu/b_1$,
is smaller in the NLTE/L model, and consequently the emergent flux is
lower.

Figure 6 displays the EUV continuum. As we have discussed in Paper I,
NLTE models produce the He II Lyman jump in emission for the hot annuli.
For cooler annuli, both LTE and NLTE models produce a strong absorption
in the He II Lyman jump. Both NLTE/C and NLTE/L models produce very similar
results.

Having computed models of vertical structure, we can
verify that neglecting Compton scattering was indeed
a legitimate approximation. To this end, we compute
the appropriate Compton $y$ parameter which specifies
whether a photon will be significantly influenced by
comptonization in traversing the medium. For non-relativistic
electrons, the Compton $y$ parameter is given by
(Rybicki \& Lightman 1979)
\begin{equation}
y = {4 k T\over m_{\rm e} c^2} \, \max(\tau_{\rm es}, \tau_{\rm es}^2)
\end{equation}
where $\tau_{\rm es}$ is the total electron-scattering optical depth
of the medium. In the case in which thermal absorption is 
not negligible, the $\tau_{\rm es}$ depends on frequency, and is defined
by (Rybicki \& Lightman 1979, their eq. 7.42)
\begin{equation}
\tau_{\rm es}(\nu)\approx \left( {\sigma_\nu/\kappa_\nu \over 1 +
\kappa_\nu/\sigma_\nu} \right)^{1/2} 
= (1-\epsilon_\nu)\, \epsilon_\nu^{-1/2}\, ,
\end{equation}
where the second equality follows from Eq. (\ref{epsnu}).
Strictly speaking, the above formulae are derived for a homogeneous medium.
We may nevertheless use them for some characteristic (averaged) values of 
structural parameters.
The Compton scattering would be most important in the hottest model.
Taking $T=T_{\rm eff}$ and the characteristic $\epsilon$ for the Lyman
continuum frequencies equal to
$10^{-2}$ (see Fig.7), we obtain for $y \approx 5 \times 10^{-3}$. Even the 
upper limit for $y$, taking the midplane values of  $T \approx 3 \times 10^5$ 
and $\epsilon \approx 2 \times 10^{-3}$ (which would give a largely
overestimated value of $y$ because the local temperature is lower and the
parameter $\epsilon$ is higher) yields $y \approx 10^{-1}$,
which is still well below unity. Therefore Compton scattering is never very
important for the models considered here.

In conclusion, there are several NLTE effects; 
for hot and cool models they 
decrease the magnitude of the Lyman discontinuity as compared to LTE
models, or to simplified NLTE/C models, while for the intermediate temperatures
they somewhat increse the Lyman jump.
We also find that for hot annuli, the
NLTE/C models are sufficient for predicting the emergent {\em continuum}
radiation, while for cooler annuli the NLTE/C models
produce a somewhat higher flux in the Lyman continuum, and spuriously strong
emissions in the Lyman lines.

We stress again that the present paper is devoted to studying a few 
representative annuli of an AGN accretion disk model. We do not aim here to
answer the important question of whether the present models will alleviate the
long-standing Lyman jump problem, namely that the theoretical models predict
a significant jump, while the observed jump is virtually non-existent.
The problem arose from early calculations by Kolykhalov \& Sunyaev (1984);
a comprehensive review of the current status was presented by Blaes (1998).
To address this problem, we have to compute theoretical models and
emergent spectra for all
annuli of a disk and integrate them using the appropriate general relativistic
transfer function (Cunningham 1975; Agol 1997). 
Also, we should take into account effects of metal lines,
which was recently found to be quite important (Hubeny \& Hubeny 1998;
Agol, Blaes, \& Hubeny 1998). We shall defer this study to a future paper
of this series.

\subsection{Effects of Viscosity}
\label{eff_visc}

The most important parameter influencing the model structure is $\alpha_0$.
We first study the effects of changing $\alpha_0$,
while keeping the remaining viscosity parameters fixed. We take 
$\zeta_0 =0$ (i.e., a constant viscosity in the inner region);
$m_{\rm d} = 10^{-2}$ (i.e., the viscosity starts to decrease with $m$ only
for $m/m_0 = 10^{-2}$; in other words, 99\% of the mass of the disk column
has a constant viscosity); and
$\zeta_1 = 2/3$.
Since the total column mass, $m_0$, is proportional to $1/\alpha_0$ --
see Eq.~(\ref{m0_alp}), we may think of effects of changing the value
of the parameter $\alpha_0$ being in fact effects of changing the disk
column mass. We stress again that our study concerns the behavior of the
individual disk annuli and its sensitivity to the adopted value of
$\alpha_0$. In reality, it is not clear whether the same value of $\alpha_0$ 
should be considered for all annuli of a given disk or not. This will only be
solved by future detailed MHD simulations similar to those of 
Stone et al. (1996).

Figure 9 shows the results for the hot annulus; the upper panel displays
the temperature, the middle panel the density, and the lower panel the emergent
flux, for the NLTE/C models for $\alpha_0$ = 0.3, 0.1, 0.03, 0.01, and 0.005.
The behavior of density is easy to understand. The total column mass,
$m_0$, is directly proportional to $1/\alpha_0$ -- see Eq.~(\ref{m0_alp}). 
The total
disk height does not depend on viscosity for radiation--pressure--dominated
disks (see Eq.~\ref{totz}); therefore the density in the inner regions is
proportional to $m_0$ -- see Eq.~(\ref{rho-m}) -- and consequently
$\rho(m \!\approx\! m_0) \propto 1/\alpha_0$. 
This is demonstrated on the middle panel of Fig.~9. 
Consequently, the thermal coupling parameter $\epsilon$ 
(which is proportional to density), is larger for smaller $\alpha_0$.

The behavior of temperature is more complicated, since it follows from 
several competing mechanisms. First, we write a simple analytic formula
derived by Hubeny (1990a) which gives a reasonably accurate estimate 
of the local temperature, viz.,
\begin{equation}
\label{tgrey}
T^4(\tau) = {3 \over 4}\, T_{\rm eff}^4\left[ \tau
\left( 1 - {\tau \over 2 \tau_{\rm tot} }\right) +  {1 \over \sqrt 3} +
{1 \over 3\,  m_0 \, \kappa_B(\tau)}\, {w(m) \over \lbar w} \right] \, ,
\end{equation}
where $\tau$ is the flux-mean optical depth, $d\tau = \chi_H dz =
(\chi_H/\rho) dm$; $\tau_{\rm tot}$ is the flux--mean optical thickness
at the midplane; and $ \kappa_B(\tau)$ is the Planck--mean opacity, defined by
\begin{equation}
\label{kappa_b}
\kappa_B = \int_0^\infty (\kappa_\nu/\rho) \, B_\nu\, d\nu/B \, ,
\end{equation}
where $B$ is the frequency-integrated Planck function, 
$B(T) = \int_0^\infty B_\nu (T) \, d\nu = (\sigma/\pi) T^4$.
This formula is easy to understand. The term $\tau + 1/\sqrt3$ in the square 
bracket is the same as for the classical LTE-grey
semi-infinite model stellar atmospheres in 
radiative equilibrium (i.e., no energy generated in the atmosphere).
The ``correction'' $(1-\tau/2\tau_{\rm tot})$ reflects the fact that the
disk has a finite total optical thickness; while the third term describes
the energy generation in the disk due to viscous dissipation. 
Notice also that the Planck--mean opacity contains the thermal absorption
coefficient, $\kappa_\nu$, while the flux--mean opacity contains the
total absorption coefficient, $\chi_\nu$. 
The fact that the dissipation term of Eq.~(\ref{tgrey}) contains the
Planck mean opacity is easy to understand, since it is the thermal
absorption, not scattering, that contributes to the global energy balance.
In the case of dominant electron scattering, the relation
between the flux--mean optical depth, $\tau$, and the column mass,
$m$, is particularly simple, $\tau \approx 0.34 \, m$ (if helium is doubly
ionized), or $\tau \approx 0.31 \, m$ (if helium is singly ionized),
as discussed in Sect.~\ref{lte-nlte}.

As follows from Eq.~(\ref{tgrey}), the temperature at the midplane is 
roughly given by
\begin{equation}
\label{tmid1}
T^4(\tau_{\rm tot}) = {3 \over 4}\,  T_{\rm eff}^4\left[ {\tau_{\rm tot} \over 2} + 
{1 \over 3\,  m_0 \, \kappa_B(\tau_{\rm tot})}\,  \right] \, ,
\end{equation}
where we assume that the disk is optically thick, $\tau_{\rm tot} \gg 1$,
and the local viscosity at the midplane roughly equal to the averaged
viscosity. Quantity $m_0 \kappa_B(\tau_{\rm tot})$ is very roughly equal to the
total Planck--mean optical depth of the disk. We write
$m_0 \kappa_B(\tau_{\rm tot}) = \lbar \epsilon \tau_{\rm tot}$. Since
$\lbar \epsilon$ is proportional to $\kappa_B/\chi_H$, it may be called the
``averaged thermal coupling parameter''. The midplane temperature then
becomes
\begin{equation}
\label{tmid2}
T^4(\tau_{\rm tot}) = {3 \over 4}\,  T_{\rm eff}^4\left( {\tau_{\rm tot} \over 2} +
{1 \over 3\, \lbar \epsilon \, \tau_{\rm tot}}\,  \right) \, ,
\end{equation}
If the electron scattering is negligible, $\lbar\epsilon \approx 1$, and the 
second term in Eq.~(\ref{tmid2}) is negligible compared to the first one in the
case of optically thick disks. In fact, the disk behaves like a normal
stellar atmosphere. However, as soon as the averaged thermal coupling
parameter becomes small, the second term in Eq.~(\ref{tmid2}) starts to
contribute. This is indeed illustrated in the upper panel of Fig.~9.
For the lowest $\alpha_0$, $\alpha_0=0.005$, the total optical depth is
very large, and consequently the central temperature is high. Increasing 
$\alpha_0$, the total optical depth decreases, and so does the central
temperature. However, with increasing $\alpha_0$ the density 
decreases, and therefore the thermal coupling parameter $\lbar \epsilon$
also decreases. 
Hence the second (viscous-heating) term of Eq.~(\ref{tmid2}) starts to make 
appreciable contribution to the central temperature,
which first levels off and then starts to increase  even if the total
optical depth decreases (see the last curve for $\alpha_0=0.3$).
Since the second term of Eq.~(\ref{tmid2}) makes roughly the same contribution
at all depths, we see a similar increase of local temperature 
at all depths. 

At the top of the disk, density decreases exponentially, and consequently
$\lbar \epsilon$ decreases faster than $w(m)/\lbar w$, which is 
assumed to be a simple
power law (recall that 
$w(m)/\lbar w \propto (m/m_0)^{2/3}$ for the models displayed
in Fig.~9). Since for the model with the largest $\alpha_0$,
$\alpha_0=0.3$, the averaged thermal coupling parameters is so small that
the second term in fact determines the temperature structure. Consequently, 
the surface temperature
begins to increase to large values (this effect is better seen in Fig.~12
where we consider a lower $\zeta_1$ and therefore a higher viscosity in the
upper layers).
However, as discussed by Hubeny (1990a),
this increase is partly spurious, because in computing the Planck--mean
opacity we have included only lines (and continua) of hydrogen and helium.
In reality, however, there is a host of other lines of light elements which
operate in the upper layers (e.g., the resonance lines of C {\sc IV},
N {\sc V}, O {\sc VI}, and higher ions of S, Ne, Fe, etc., for even higher
temperatures).  Including these lines will increase the Planck--mean opacity
significantly, which physically means that we are including efficient 
mechanisms which are able to radiate the dissipated energy away.

The lower panel of Fig.~9 shows the continuum flux for all
models. The Lyman jump varies from being almost non-existent for the
low--$\alpha_0$ models, 
to a weak absorption jump for the high--$\alpha_0$ models.
As explained above, the appearance of the Lyman jump is a result of a 
competition of two mechanisms: the overpopulation of the ground state of
hydrogen, which decreases the source function and therefore the emergent
flux in the Lyman continuum; and a higher thermal coupling parameter which
causes the mean intensity in the Lyman continuum frequencies to be more coupled
to the thermal source function, and therefore causes the emergent flux to
be higher. In the present case, going to higher $\alpha_0$ decreases
density in the inner layers, therefore causing the hydrogen ground state to
be more and more overpopulated, and since this is a dominant effect, the
flux in the Lyman continuum decreases with increasing $\alpha_0$, so the Lyman
jump is driven to a somewhat stronger absorption.

In contrast, the magnitude of the He II Lyman jump decreases with increasing
$\alpha_0$, from being a very conspicuous emission jump at $\alpha_0\lta 0.03$ 
to a modest emission jump at $\alpha_0=0.3$. This is explained by the fact that
the strong jump at $\alpha_0\lta 0.03$ is essentially caused by the Schuster
mechanism discussed above. 
When going to higher $\alpha_0$ the density decreases, and
the electron scattering is more and more important; the jump thus becomes
weaker. However, it is important to realize that the number of He II 
ionizing photons remains roughly the same for all models, since the flux
at high frequencies ($\nu \gta 1.7 \times 10^{16}$ Hz) is higher for
higher $\alpha_0$; this is a consequence of higher temperature in
the continuum--forming region -- see the upper panel of Fig.~9.

Figure 10 presents an analogous comparison of models with various values of
$\alpha_0$ for a ``cool'' annulus, $T_{\rm eff} = 18,000$ K.
The behavior of models is similar to that of the hot annulus, although
the effects of viscosity are generally weaker. The total
column mass and the inner density are proportional to $1/\alpha_0$, as in the
previous case. The central temperature now decreases monotonically with
increasing $\alpha_0$ because the electron scattering is not dominant and 
therefore the second term in Eq.~(\ref{tmid2}) is not very important. 
The Lyman jump varies from a weak emission for high--$\alpha_0$ models to
a weak absorption for low--$\alpha_0$ models.  This follows from the
fact that for increasing $\alpha_0$ the density decreases, so the magnitude of
NLTE effects increases. In the present case, the ground state of hydrogen 
becomes more underpopulated, and consequently the flux in the Lyman continuum
increases.

Next, we examine the effects of changing the other viscosity parameters.
In Fig.~11, we display the models of the hot annulus computed 
for fixed $\alpha_0$ ($\alpha_0=0.1$)
and the power law exponents ($\zeta_0=0$ and $\zeta_1=2/3$), and
for several values  of the division mass, $m_{\rm d}$. 
The division mass is varied from $m_{\rm d} = 1$ (i.e., no inner region of
constant viscosity), to $m_{\rm d} = 0.3,\, 0.1,\,  0.03$, and $0.01$.
As expected, the only interesting effect is the behavior of density in
the inner layers, which is governed by Eq.~(\ref{rho-m}).
The central temperature is the largest for $m_{\rm d} = 1$ because the central
density, and thus the thermal coupling parameter, are lowest. The emergent
flux is only weakly influenced by changing $m_{\rm d}$; a modest change of the
flux is seen in the hydrogen and He II Lyman continuum which are the most 
sensitive to the thermal coupling parameter.

Finally, Fig.~12 displays the effect of changing $\zeta_1$,
$\zeta_1 = 1,\,  2/3,\,  1/2$, and $1/3$, for $\alpha_0=0.1$, 
$m_{\rm d} = 1$, for
the hot annulus ($r=2$). For $\zeta_1 \gta 1/2$, the only appreciable 
effects are seen in the behavior of density in inner layers. 
The overall behavior of the models
is easily explained by the same considerations as above.
For the model with the lowest $\zeta_1$, $\zeta_1 = 1/3$, 
the local viscosity in the upper
layers is so high that we see the effects of temperature runaway at 
$m \approx 5 \times 10^{-3}$. Decreasing the value of $\zeta_1$ still would
increase this instability. 
However, as discussed above, we cannot study this effect
with simple H-He models, and we leave this problem to a future paper.

%
%

\subsection{Comparison with the Local $\alpha$--viscosity Approach}

As discussed in Sect.~2.2, we consider the depth-dependent kinematic
viscosity as a step-wise power law function given by 
Eqs.~(\ref{visc1}) -- (\ref{w1}). In contrast, the local
$\alpha$--prescriptions, e.g., the variant 
suggested by D\"orrer et al. (1996), considers
the kinematic viscosity in the form given by Eqs.~(\ref{visc_alpha1}), with
the turbulent velocity given by
\begin{equation}
\label{dorrer_vtb}
v_{\rm turb}=c_{\rm s}\,{\tau+\sqrt{P_{\rm gas}/P} \over \tau + 1} \, ,
\end{equation}
This approach essentially
considers the viscosity being proportional to the gas pressure at
low optical depths, while being proportional to the radiation pressure at 
large depths; $\tau$ is the Rosseland mean optical depth.

We can compute an ``effective $\alpha$--parameter''
in such a way that the two prescriptions give the same value of local
viscosity,
\begin{equation}
\alpha_{\rm eff}(m) = {w(m) \over h \, v_{\rm turb}(m)}\, .
\end{equation}
Since our parameterization does not take viscosity to be proportional
to the local turbulent velocity, the resulting $\alpha_{\rm eff}$ will be
depth-dependent.

In Fig.~13, we plot $\alpha_{\rm eff}$ for the hot annulus ($r=2$),
and for various viscosity parameters. We see that the values of 
$\alpha_{\rm eff}$
for our standard model, $\alpha_0=0.1$, $\zeta_0 = 0$, $\zeta_1 = 2/3$, 
and $m_{\rm d}/m_0 = 0.01$ are located in a reasonable range of
$0.02 \lta \alpha_{\rm eff} \lta 0.5$. 
The behavior of $\alpha_{\rm eff}$ as a function of depth is easily
understood. In the inner layers, 
$v_{\rm turb} = (P/\rho)^{1/2} \approx (P_{\rm rad}/\rho)^{1/2}$. Since the
optical depth is large, the radiation pressure can be approximated by
the thermodynamic equilibrium form, $P_{\rm rad} \propto T^4$. In the
model with $m_{\rm d}/m_0 = 0.01$ density is roughly constant for
$m_{\rm d} \lta m \leq m_0$. Consequently, $v_{\rm turb} \propto T^2$.
To first order, $T^4 \propto \tau \propto m$ (see Eq.~\ref{tgrey}), so that
finally $v_{\rm turb} \propto m^{1/2}$, and $\alpha_{\rm eff} \propto m^{-1/2}$
for $m > m_{\rm d}$. This is indeed seen in the upper and lower panels of
Fig.~13; $\alpha_{\rm eff}$ decreases with $m$ somewhat slower than 
$m^{-1/2}$ because the temperature increases slower than $T^4 \propto m$ --
in fact, $T^4 \propto (m - m^2/2m_0)$.

For models with a single power law viscosity (i.e., $m_{\rm d} = m_0$),
the behavior of $\alpha_{\rm eff}$ in the inner layers can also be easily
understood. Here $w(m) \propto m^\zeta$ (we write $\zeta$ instead of $\zeta_1$
to simplify the notation), so the density varies as $\rho(m) \propto 1/w(m)
\propto m^{-\zeta}$. The radiation pressure scales, as discussed above,
as $P_{\rm rad} \propto m$. Consequently, 
$v_{\rm turb} \propto m^{(\zeta+1)/2}$,
and thus $\alpha_{\rm eff} \propto m^{(\zeta-1)/2}$. This is demonstrated
in the middle panel of Fig.~13; for instance, we see that for $\zeta=1$,  
$\alpha_{\rm eff}$
is indeed almost constant in the inner layers. In the outer layers,
the density decreases outward faster than $w(m)$, so $\alpha_{\rm eff}$
decreases faster than $m^{(\zeta-1)/2}$.

In the gas--pressure--dominated regions, 
$v_{\rm turb} = (P_{\rm gas}/\rho)^{1/2}$,
so we have $v_{\rm turb} \propto T$, and thus 
$\alpha_{\rm eff} \propto w(m)/T(m)$. In the outer layers, the temperature
is roughly constant with $m$, so that $\alpha_{\rm eff} \propto w(m) \propto
m^{\zeta_1}$. Again, this is clearly seen in Fig.~13.

The important point to realize is that although the local value of 
$\alpha_{\rm eff}$ varies significantly, its influence on the disk
structure is rather small. Consider for instance the middle panel of
Fig.~13, and compare it to Fig.~12, which displays the same models.
Although $\alpha_{\rm eff}$ differs by several orders of magnitude, the
disk structure is hardly affected. 
From the point of view of constructing detailed models of vertical structure
of AGN disks this is a good news, because the most uncertain part of physics
-- the viscous dissipation -- has relatively small effect on the computed
structure. However, we should bear in mind that this study was limited to 
considering a ``well-behaved'' viscosity which smoothly decreases towards the 
disk surface. When one
assumes, for instance, that the viscous dissipation is concentrated mostly
in the outer layers (e.g., Sincell \& Krolik 1997), the overall vertical
structure may be significantly different. We plan to study such non-standard 
models in future papers of this series.

%
%

\section{CONCLUSIONS}

We have calculated several representative models of vertical structure
of an accretion disk around a supermassive Kerr black hole. The interaction
of radiation and matter is treated self-consistently, taking into account
departures from LTE for calculating both the disk structure and the
radiation field. The viscosity is parameterized through the 
parameter $\alpha_0$ that describes the vertically averaged viscous stress, 
and two power--law exponents $\zeta_0$ and $\zeta_1$, and the division point
$m_{\rm d}$ between these two forms.
The disk structure and emergent radiation is 
sensitive mainly to the values of $\alpha_0$, while the other parameters
influence the disk structure to a much lesser extent. However, although
the detailed shape of the predicted spectrum is sensitive to adopted
$\alpha_0$, the overall appearance of the spectrum is quite
similar in all cases. 

We have shown that effects of departures from LTE are very important for
determining the disk structure and emergent radiation, particularly for
hot and electron--scattering dominated disks. We have shown that at least
for the disk parameters studied in this paper, NLTE effects typically tend to
diminish the value of the Lyman jump; in hot models they suppress the
Schuster mechanism by which the LTE models produce a strong emission jump,
and in cooler models they increase the flux in the Lyman continuum due
to an underpopulation of the hydrogen ground state. Also, we have shown that
relaxing the approximation of detailed radiative balance in the hydrogen and 
helium lines (i.e., computing the so-called
NLTE/L models) changes the predicted line profiles significantly, but
otherwise does not yield significant changes in computed vertical structure
or emergent continuum flux. This result shows that for estimating
the {\em continuum} radiation of AGN disks composed of hydrogen and helium,
the NLTE/C models provide a satisfactory approximation.

So far, we have limited our analysis to a simple H-He chemical composition.
A preliminary study (Hubeny \& Hubeny 1998) indicates that the effects of
numerous metal lines on the prediced spectral energy distribution of
AGN disks may be quite significant. However, that study used a simplified
approach in which the vertical structure was fixed by a H-He model,
while the line opacity was taken into account only in the spectrum synthesis,
assuming LTE source function in metal lines. Such an approach is
inconsistent, since, first, the metal lines may change the disk vertical 
structure (the so-called metal line blanketing effects, long known form the 
theory of classical stellar atmospheres) and, second, the source function in 
metal lines may depart significantly from the LTE value. We will therefore
need to construct self-consistent, fully metal-line-blanketed
models of vertical structure of AGN disks, taking into account effects
of literally millions of spectral lines in NLTE. A work of this project
is under way, and will be reported in a future paper of this series.

The results presented here do not indicate any fatal flaw of the AGN accretion
disk paradigm. In contrast, they show that one of the previous critical 
arguments against the accretion disk paradigm, the magnitude of the Lyman jump,
essentially disappears when increasing a degree of realism of the
modeling procedure by relaxing previous simplifying approximations, in
particular the local thermodynamic equilibrium and a simplified vertical disk
structure.
However, this study has concentrated on only one aspect
of the problem, the spectral energy distribution in the optical, UV, and
EUV region. Many questions, such as the overall spectral energy distribution
of the whole disk, the effects of external irradiation, the continuum 
polarization, etc., remain to be explored in detail. This is exactly what we 
intend to do in future papers of this series.

%
%

\acknowledgments
This work was supported in part by NASA grant NAGW-3834 and by the
HST/STIS project funds.
We thank Eric Agol, Omer Blaes, and Julian Krolik for valuable discussions and
very helpful comments on the manuscript.
%
%

\clearpage

\begin{figure}
\epsscale{0.7}
\plotone{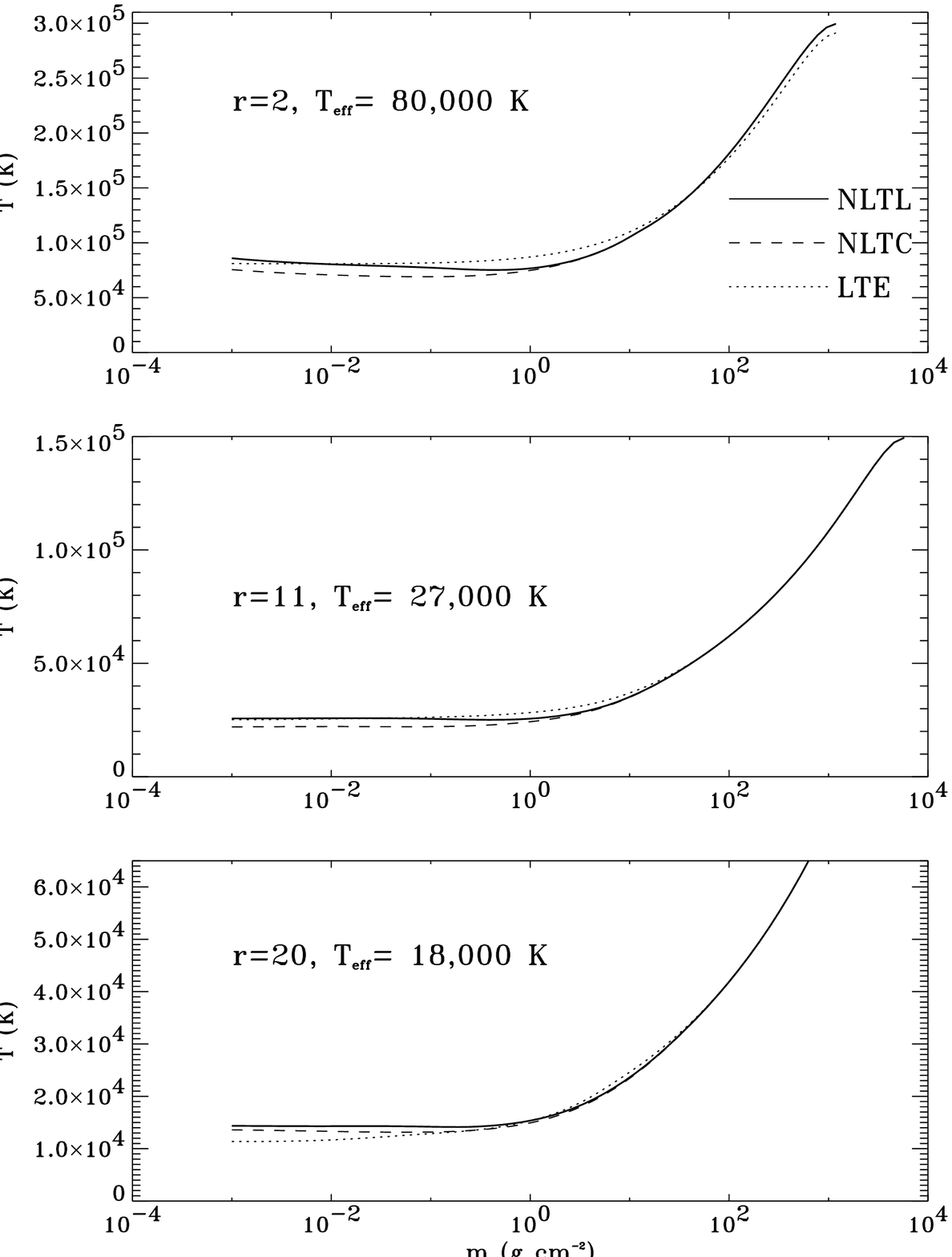}
\caption{
Temperature as a function of depth for three annuli at $r=2$ -- upper panel;
$r=11$ -- middle panel; and $r=20$ -- lower panel; for a disk model
with the mass of the black hole, $M=2 \times 10^9 M_\odot$, the mass 
accretion rate $\dot M = 1\, M_\odot$/yr, and the maximum stable rotation, 
$a/M= 0.998$. The viscosity parameters are taken $\alpha_0=0.1$, $\zeta_0=1$,
$\zeta_1 = 2/3$, and $m_{\rm d} =0.01$. For all annuli, the thick line
is the NLTE/L model, the dashed line the NLTE/C (i.e., NLTE with continua
only) model, and the dotted line is the LTE model.
}
\end{figure}

\begin{figure}
\plotone{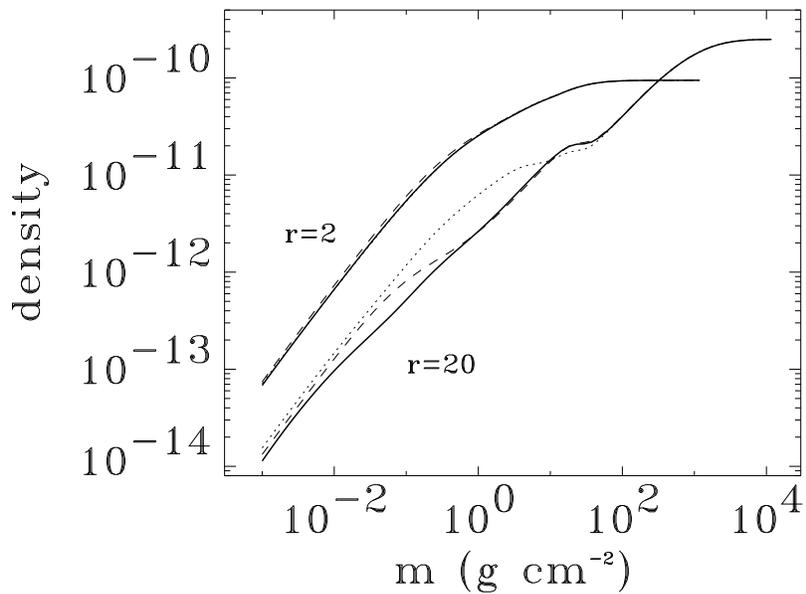}
\caption{
Mass density (in g cm$^{-3}$) for the model annulus at $r=2$ -- the upper
curves, and for $r=20$ -- the lower curves; for the same models as
displayed in Fig.~1. We did not show the $r=11$ model because its behavior
is quite analogous to that of the $r=20$ model.
}
\end{figure}

\begin{figure}
\plotone{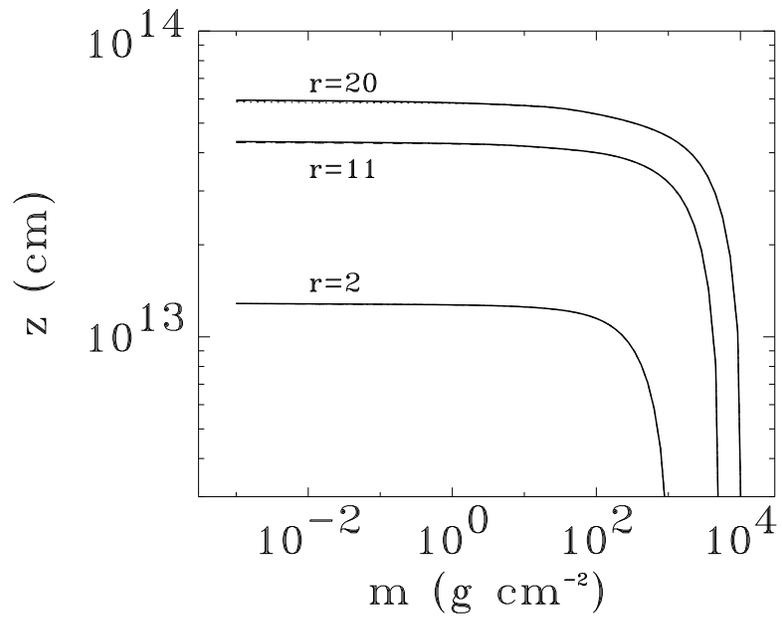}
\caption{
Vertical distance from the disk plane as a function of column mass
for the model disk annulus at $r=2$  (lower curves);
$r=11$ (middle curves); and $r=20$ (upper curves), for the same models
as displayed in Fig.~1. Notice that NLTE effects upon the $z$ vs. $m$
relation are quite small.
}
\end{figure}

\begin{figure}
\epsscale{0.8}
\plotone{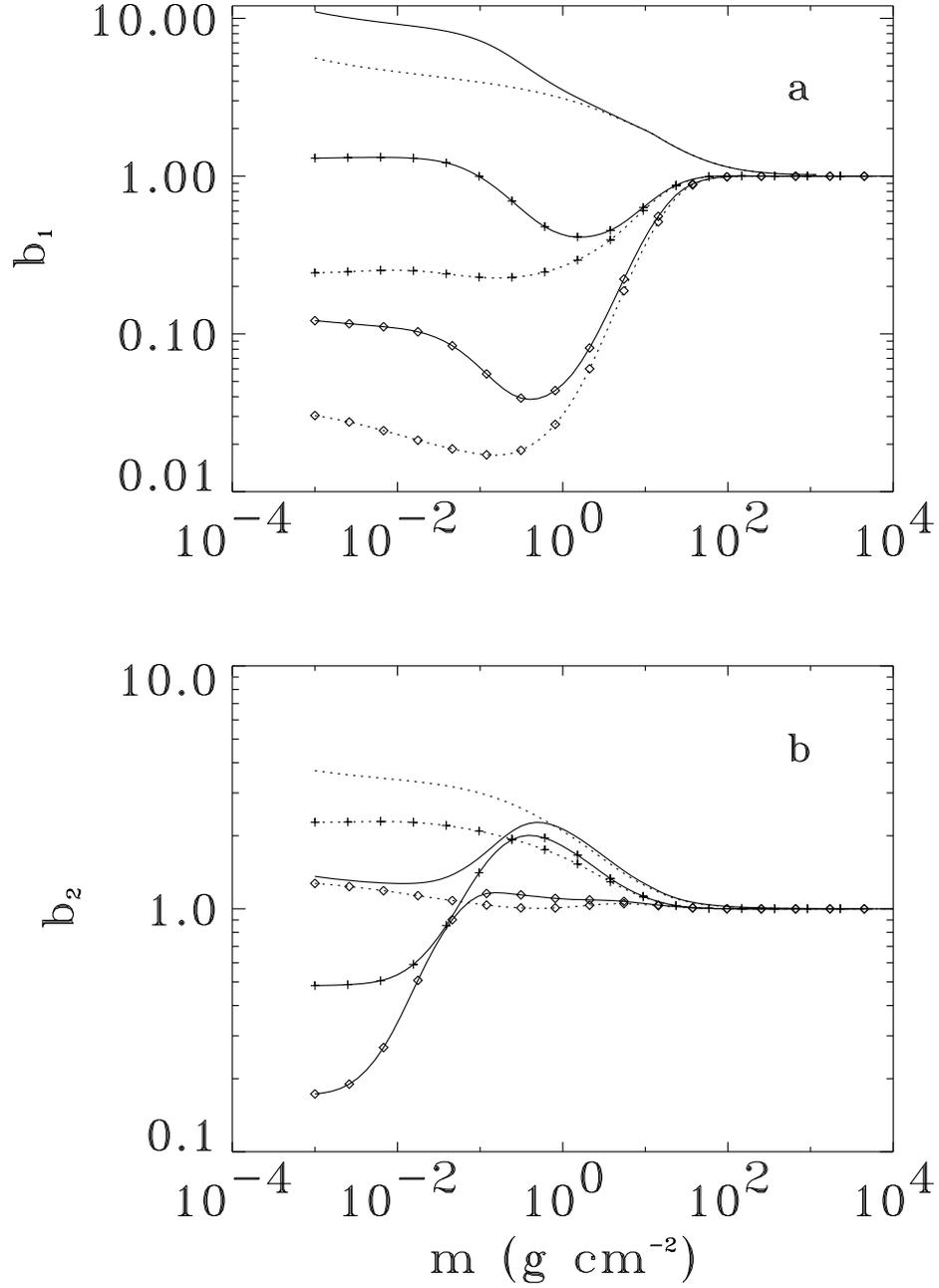}
\caption{
NLTE departure coefficients for the hydrogen ground level ($a$),
and the $n=2$ level ($b$); for the NLTE/C models (dotted lines),
and NLTE/L models (full lines). The lines without additional symbols
correspond to the model annulus at $r=2$; the lines with additional ``+''
signs to the $r=11$ models, and with diamonds to $r=20$ models.
}
\end{figure}

\begin{figure}
\epsscale{0.8}
\plotone{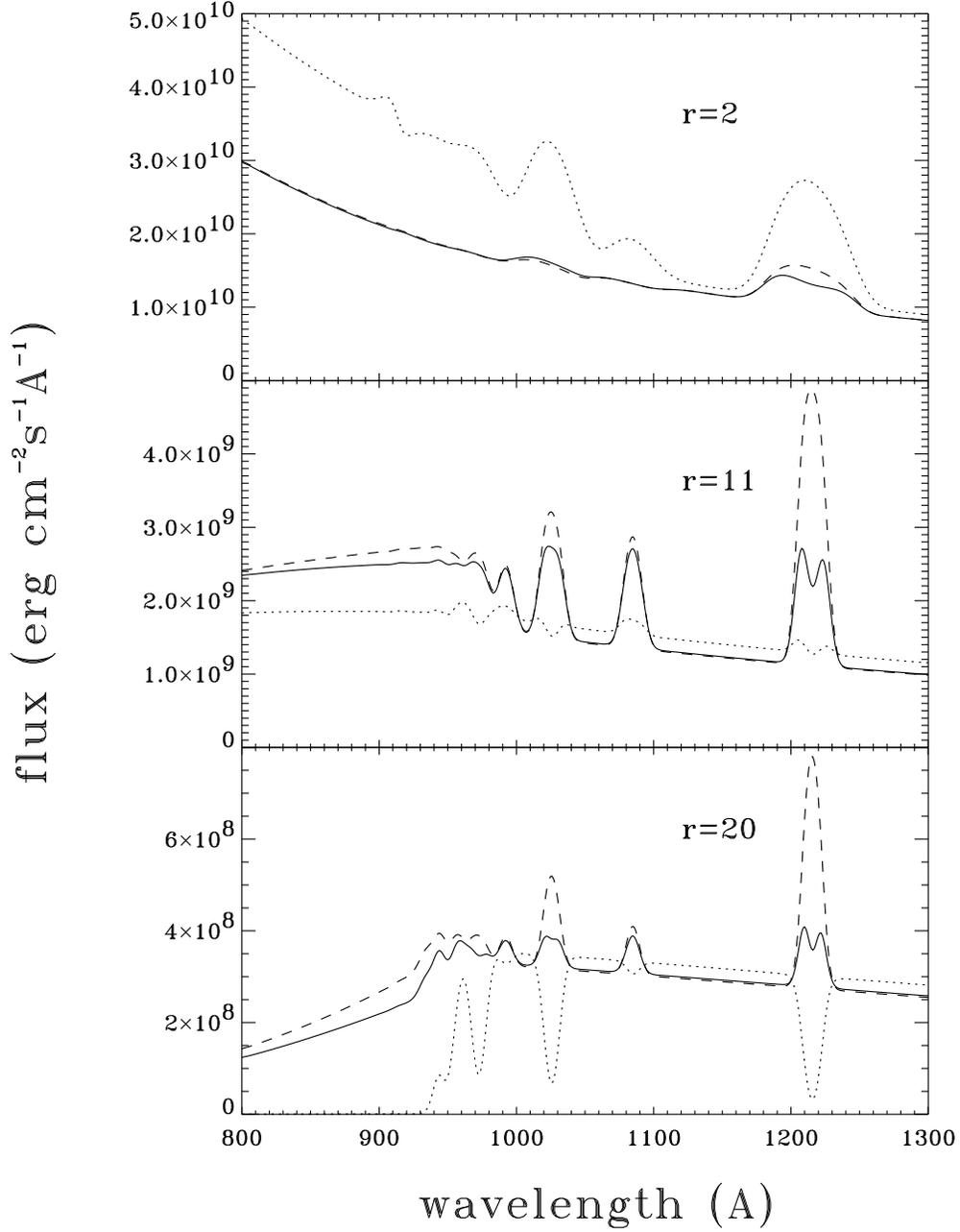}
\caption{
A comparison of emergent flux in the region around the Lyman discontinuity
for the NLTE/L model (full line); NLTE/C
model (dashed line), and LTE model (dotted line) of the model disk
annulus at $r=2$ (upper panel), $r=11$ (middle panel), and $r=20$ (lower
panel).
}
\end{figure}

\begin{figure}
\epsscale{0.8}
\plotone{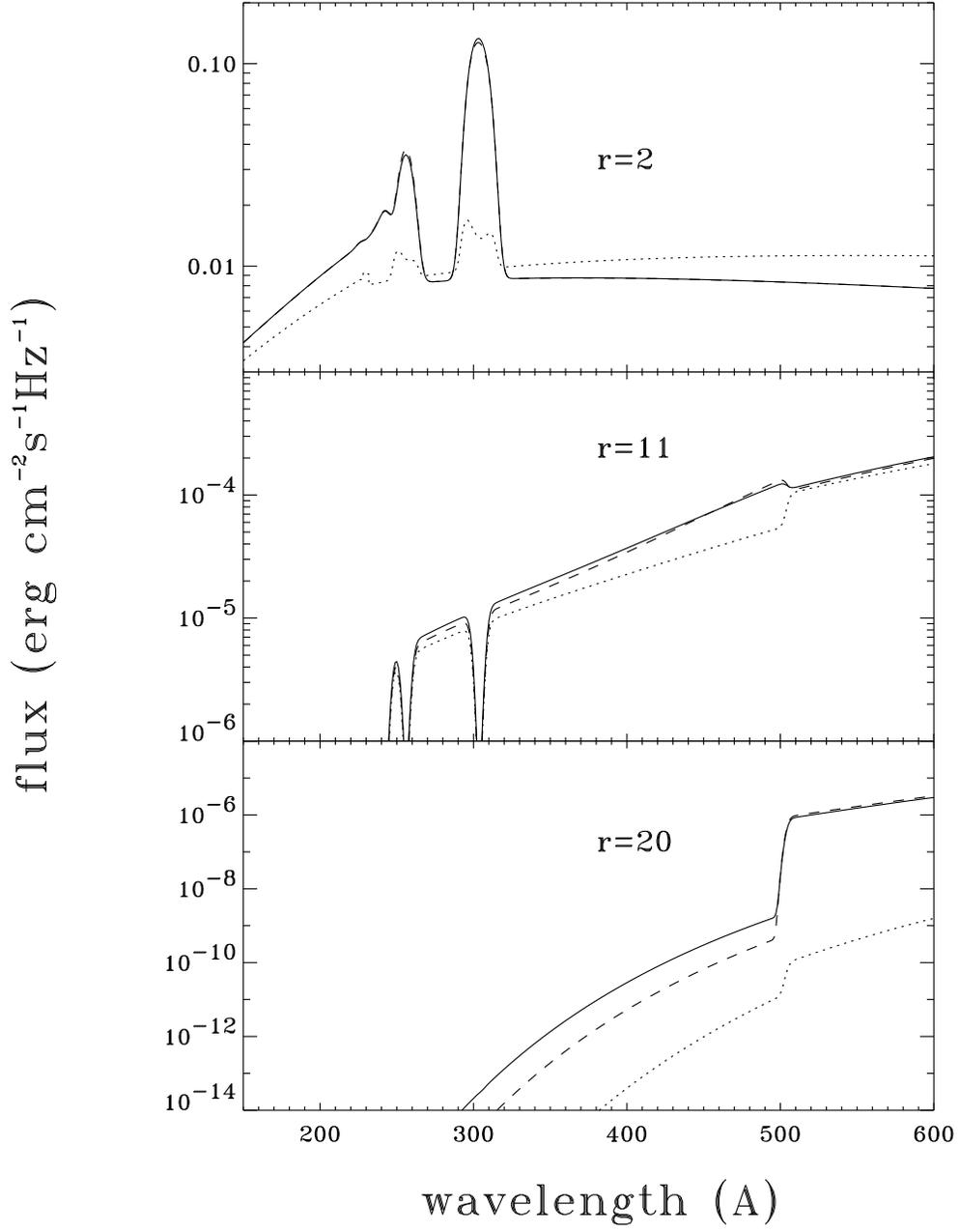}
\caption{
The same as in Fig.~5 for the EUV spectrum at the region of He II Lyman
discontinuity ($\lambda = 227$ \AA), and the He I ground--state
discontinuity ($\lambda = 504$ \AA).
}
\end{figure}

\begin{figure}
\plotone{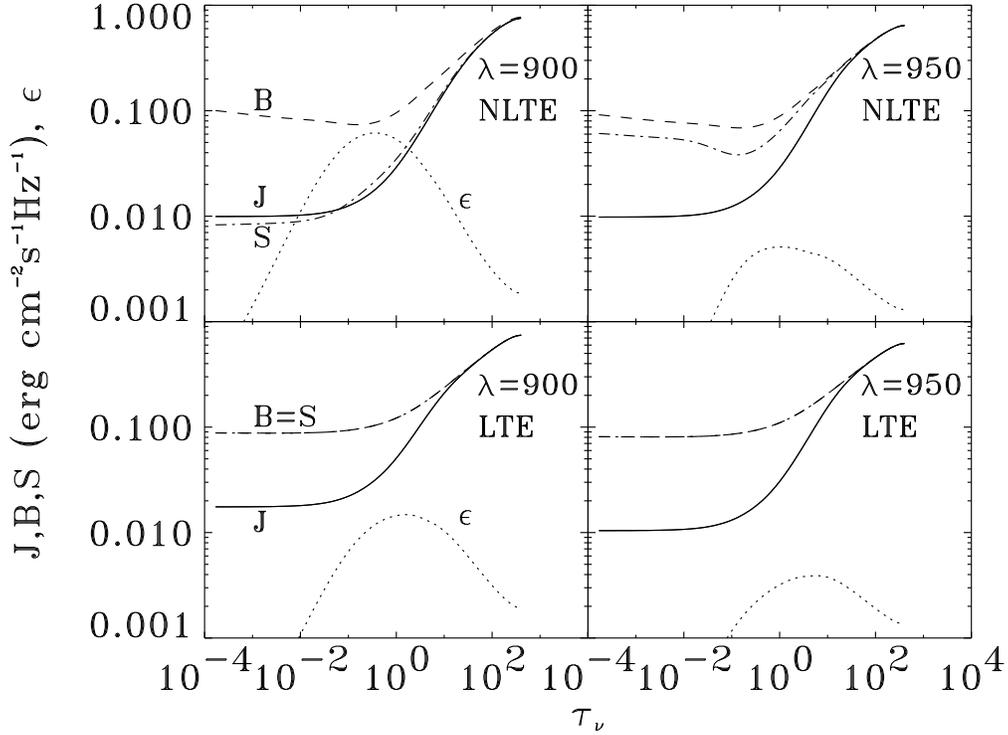}
\caption{
Plot of the mean intensity of radiation (full line, labeled $J$), the
Planck function (dashed line, labeled $B$), the thermal source function
(dot-dashed line, labeled $S$), and the thermal coupling parameter
$\epsilon$ (dotted line), as a function of the monochromatic optical
depth, for the model annulus at $r=2$. The two upper panels are NLTE models;
the two lower panels are LTE models. The left panels display the values 
for the blue side of the Lyman discontinuity ($\lambda = 900$ \AA), 
and the right panels display the values for the red side of the
Lyman discontinuity ($\lambda = 950$ \AA).
}
\end{figure}

\begin{figure}
\plotone{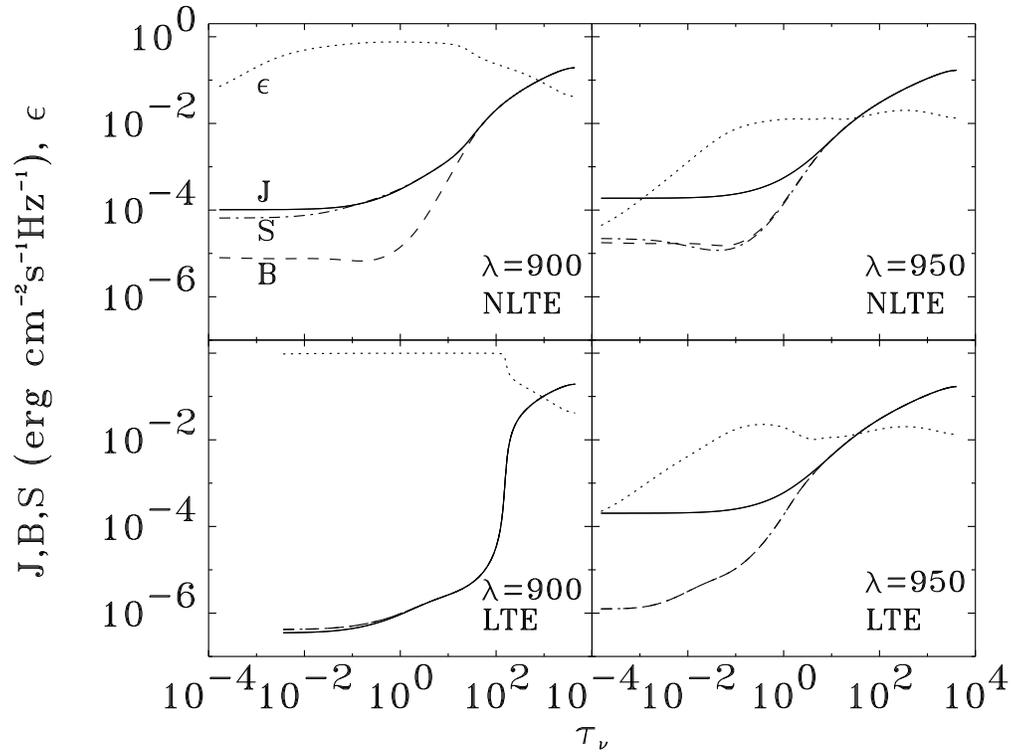}
\caption{
The same as in Fig.~7, but for the model annulus at $r=20$.
}
\end{figure}

\begin{figure}
\epsscale{0.7}
\plotone{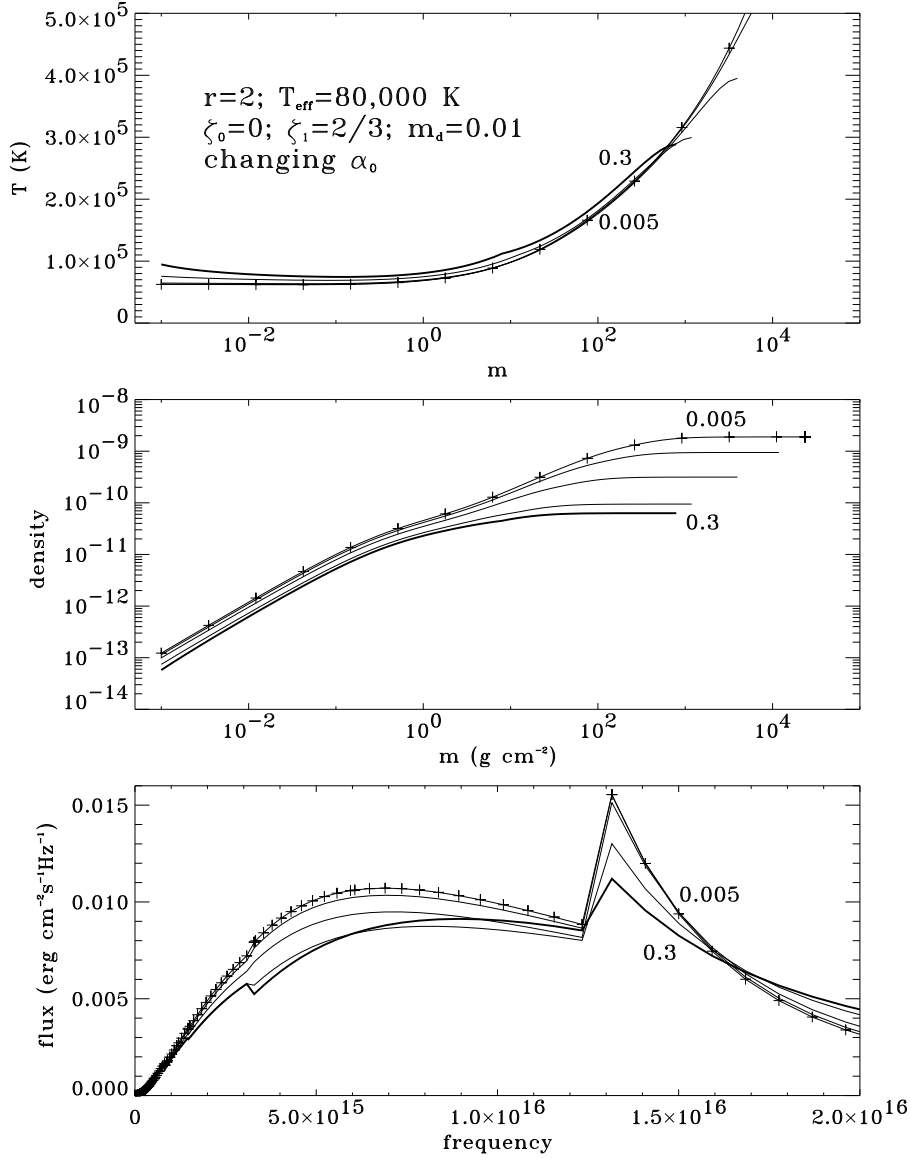}
\caption{
The effect of $\alpha_0$ on the vertical structure and
emergent flux for the model annulus at $r=2$. Other basic parameters are
the same as for the models displayed in Fig.~1, namely $\zeta_0=1$,
$\zeta_1 = 1$, and $m_{\rm d} =0.01$. Upper panel: temperature;
middle panel: mass density; and lower panel: emergent flux. The thick
line corresponds to the highest $\alpha_0$, $\alpha_0=0.3$, the line with 
additional ``+'' signs to the lowest one, $\alpha_0=0.005$,
and the curves in between to $\alpha_0=0.1,\, 0.03$, and 0.01.
The curves for the two extreme values are labeled by
the values of $\alpha_0$.
}
\end{figure}

\begin{figure}
\epsscale{0.75}
\plotone{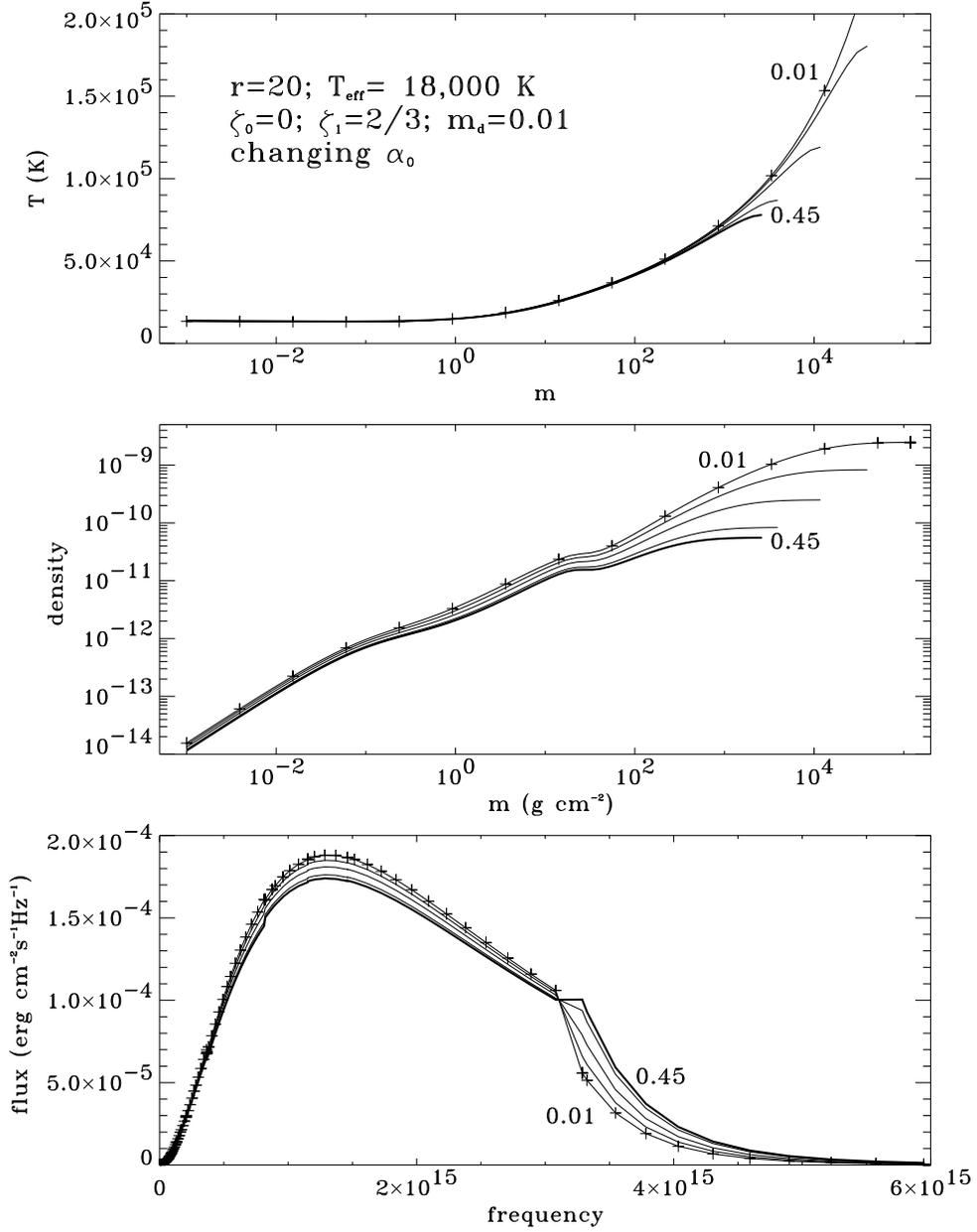}
\caption{
Analogous to Fig.~7, for the model annulus at $r=20$; the adopted $\alpha_0$ 
values are 0.45 (thick line), 0.3, 0.1, 0.03, and 0.01
(the line with additional ``+'' signs).
}
\end{figure}

\begin{figure}
\epsscale{0.7}
\plotone{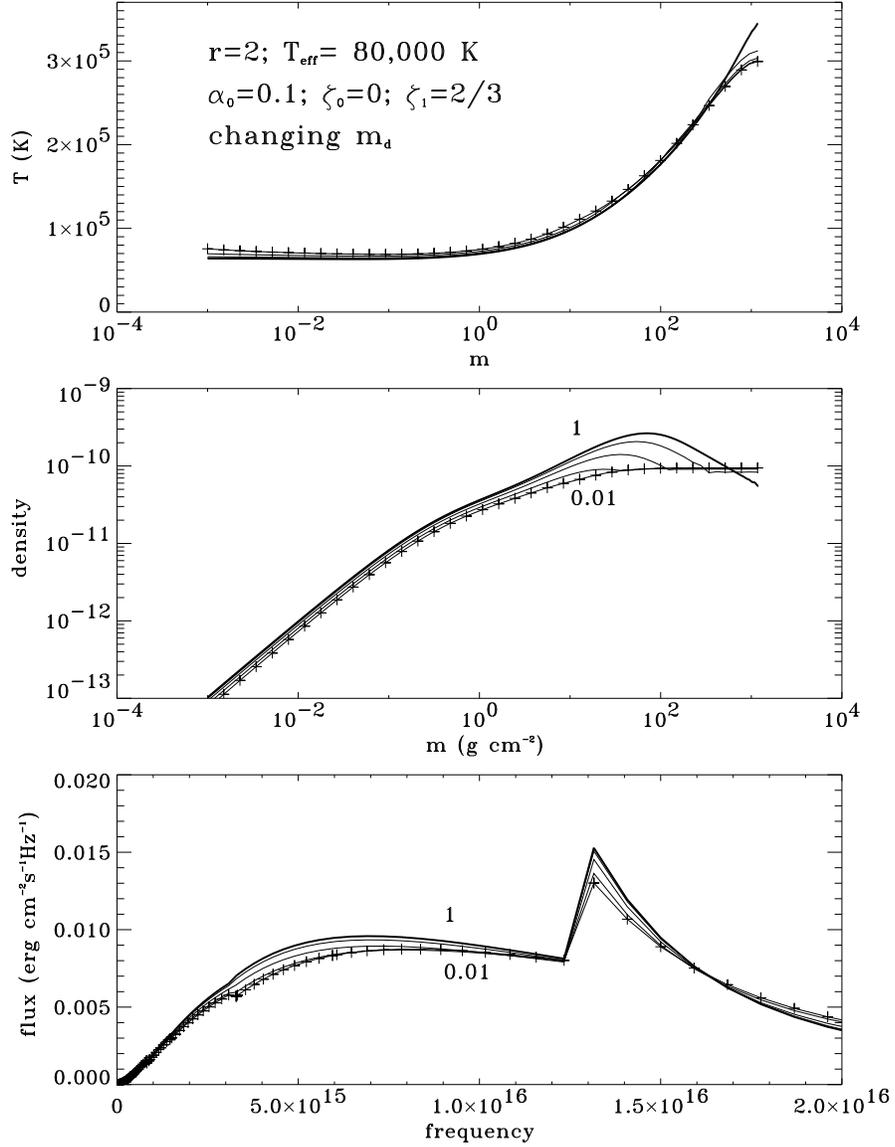}
\caption{
The effect of $m_{\rm d}$ (the division point between two different power 
laws of viscosity parameterization) on the vertical structure and
emergent flux for the model annulus at $r=2$. Other basic parameters are
the same as for the models displayed in Fig.~1, namely $\zeta_0=1$,
$\zeta_1 = 2/3$, and $\alpha_0=0.1$. The panels are arranged as in Fig.~7.
The adopted values of $m_{\rm d}$ are: 1 (thick line); 0.3; 0.1; 0.03;
and 0.01 (the line with additional ``+'' signs). 
The curves for the two extreme values are labeled by
the values of $m_{\rm d}$.
}
\end{figure}

\begin{figure}
\epsscale{0.7}
\plotone{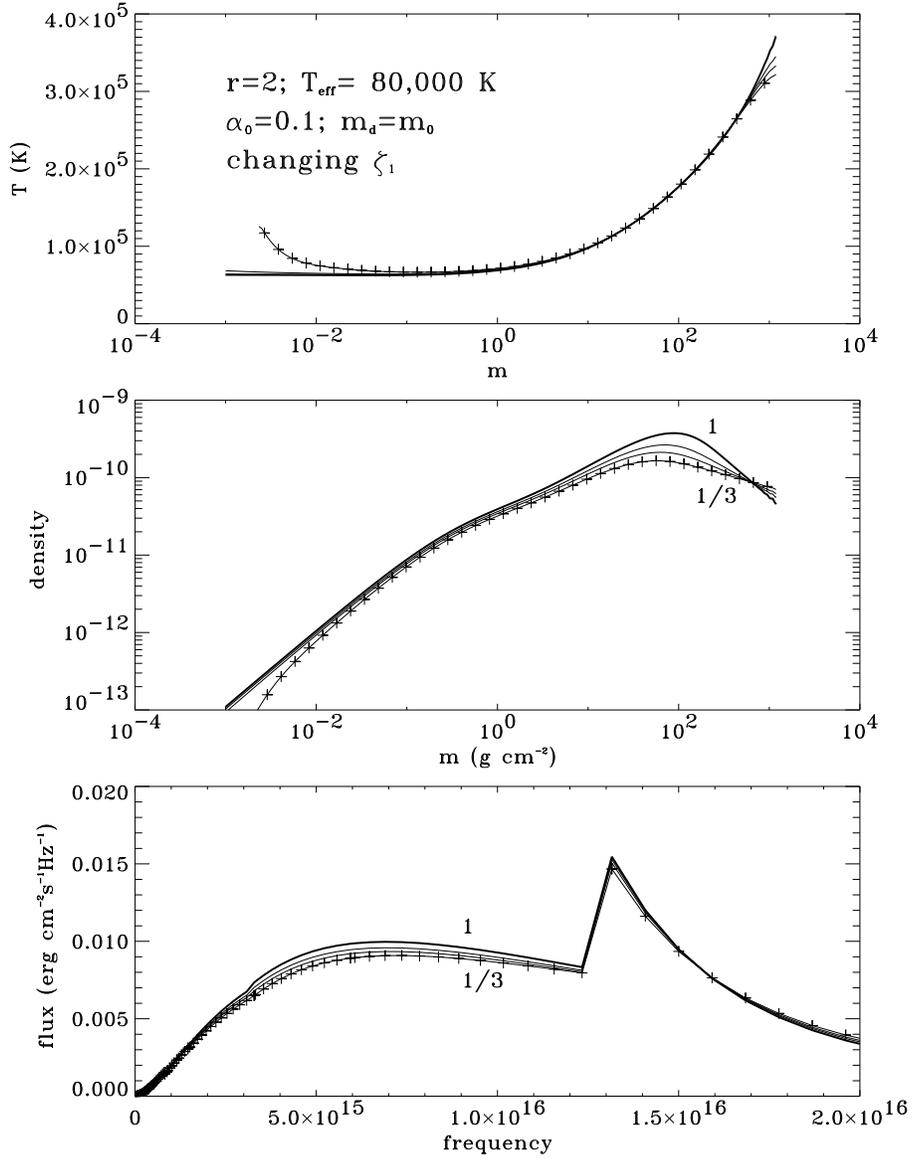}
\caption{
The effect of the power law exponent $\zeta_1$  on the vertical structure and
emergent flux for the model annulus at $r=2$. Other basic parameters are
$\alpha_0=0.1$, and $m_{\rm d}=1$ (i.e., no inner region of constant viscosity). 
The panels are arranged as in Fig.~7.
The adopted values of $\zeta_1$ are: 1 (thick line); 2/3; 1/2; 
and 1/3 (the line with additional ``+'' signs). 
The curves for the two extreme values are labeled by
the values of $\zeta_1$.
}
\end{figure}

\begin{figure}
\epsscale{0.6}
\plotone{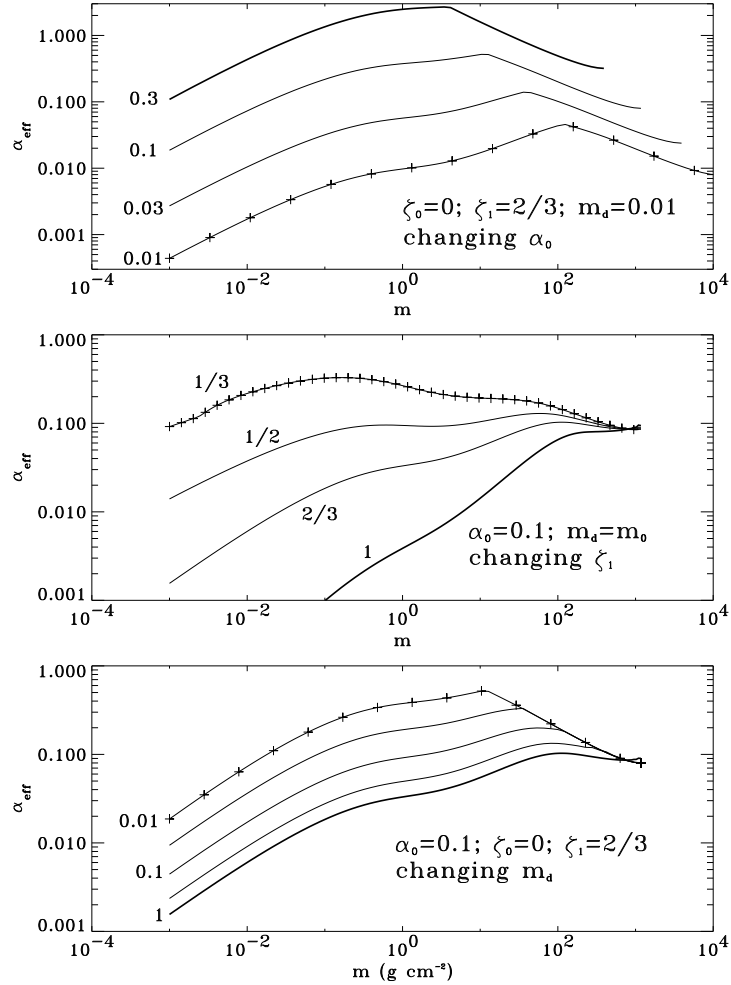}
\caption{
Plot of effective $\alpha$--viscosity parameter that would yield the
same value of kinematic viscosity in our models and in the 
$\alpha$--parameterization of viscosity suggested by D\"orrer et al. (1996). 
Upper panel --
$\alpha_{\rm eff}$ for the models displayed in Fig.~7, i.e., the models of
the $r=2$ annulus with  $\zeta_0=0$, $\zeta_1 = 2/3$, and $m_{\rm d} =0.01$,
and with various values of $\alpha_0$. 
The adopted values of $\alpha_0$ are (from top to bottom):
0.3 (thick line), 0.1, 0.03 and 0.01
(the line with additional ``+'' signs).
Middle panel --
$\alpha_{\rm eff}$ for the models displayed in Fig.~10, i.e., the models of
the $r=2$ annulus with  $\alpha_0=0.1$, $m_{\rm d}=1$,
and with various values of $\zeta_1$. 
The adopted values of $\zeta_1$ are (from bottom to top): 1 (thick line); 
2/3; 1/2; and 1/3 (the line with additional ``+'' signs). 
Lower panel --
$\alpha_{\rm eff}$ for the models displayed in Fig.~9, i.e., the models of
the $r=2$ annulus with $\alpha_0=0.1$, $\zeta_0=0$, 
$\zeta_1 = 2/3$, and with various values of the division depth $m_{\rm d}$.
The adopted values of $m_{\rm d}$ are (from bottom to top): 1 (thick line); 
0.3; 0.1; 0.03; and 0.01 (the line with additional ``+'' signs). 
}
\end{figure}

\end{document}